\documentclass[twocolumn]{aastex631}

\usepackage{color}

\usepackage{enumitem}
\usepackage{amsmath}
\usepackage{gensymb}
\usepackage{soul}

\newcommand{\be}{\begin{equation}}
\newcommand{\ee}{\end{equation}}

\newcommand{\lcdm}{\ensuremath{\Lambda\mathrm{CDM}}}

\providecommand{\sorthelp}[1]{}

\begin{document}

\title{Revisiting the $A_L$ Lensing Anomaly in Planck 2018 Temperature Data}

\shorttitle{Revisiting the Planck Lensing Anomaly}
\shortauthors{G.~E.~Addison et al.}

\author[0000-0002-2147-2248]{Graeme E. Addison}
\correspondingauthor{Graeme E. Addison}
\email{gaddison@jhu.edu}
\affiliation{The William H. Miller III Department of Physics and Astronomy, Johns Hopkins University, 3400 N. Charles St., Baltimore, MD 21218-2686}

\author[0000-0001-8839-7206]{Charles L. Bennett}
\affiliation{The William H. Miller III Department of Physics and Astronomy, Johns Hopkins University, 3400 N. Charles St., Baltimore, MD 21218-2686}

\author[0000-0002-1760-0868]{Mark Halpern}
\affiliation{Department of Physics and Astronomy, University of British Columbia, Vancouver, BC, V6T 1Z1, Canada}

\author[0000-0002-4241-8320]{Gary Hinshaw}
\affiliation{Department of Physics and Astronomy, University of British Columbia, Vancouver, BC, V6T 1Z1, Canada}

\author[0000-0003-3017-3474]{Janet L. Weiland}
\affiliation{The William H. Miller III Department of Physics and Astronomy, Johns Hopkins University, 3400 N. Charles St., Baltimore, MD 21218-2686}

\begin{abstract}

\noindent
We revisit the lensing anomaly in the Planck 2018 temperature (TT) data and examine its robustness to frequency selection and additional sky masking. Our main findings are: (1) The phenomenological lensing amplitude parameter, $A_L$, varies with ecliptic latitude, with a $2.9\sigma$ preference for $A_L>1$ near the ecliptic, and $1.0\sigma$ preference near the ecliptic poles, compared to $2.5\sigma$ on the original masks. This behavior is largely or solely from 217~GHz and suggestive of some non-random effect given the Planck scan strategy. (2) The 217~GHz TT data also show a stronger preference for $A_L>1$ than the lower frequencies. The shifts in $A_L$ from 217~GHz with additional Galactic dust masking are too large to be explained solely by statistical fluctuations, indicating some connection with the foreground treatment. Overall, the Planck $A_L$ anomaly does not have a single simple cause. Removing the 217~GHz TT data leaves a $1.8\sigma$ preference for $A_L>1$. The low-multipole ($\ell<30$) TT data contribute to the preference for $A_L>1$ through correlations with \lcdm\ parameters. The 100 and 143~GHz data at $\ell\geq30$ prefer $A_L>1$ at $1.3\sigma$, and this appears robust to the masking tests we performed. The lensing anomaly may impact fits to alternative cosmological models. Marginalizing over $A_L$, optionally applied only to Planck TT spectra, can check this. Models proposed to address cosmological tensions should be robust to removal of the Planck 217~GHz TT data.
\end{abstract}

\keywords{\href{http://astrothesaurus.org/uat/322}{Cosmic microwave background radiation (322)}; 
\href{http://astrothesaurus.org/uat/1146}{Observational cosmology (1146)}}

\section{Introduction}
\label{sec:intro}

The Planck cosmic microwave background (CMB) observations are a crucial dataset in modern cosmology, currently providing the most precise constraints in the standard \lcdm\ model, as well as tightly constraining many alternative models or \lcdm\ extensions \citep{planck2016-l06}. 

While many other measurements are consistent with the Planck \lcdm\ parameter determination, some exhibit moderate to severe tension. Most notably, the Cepheid-supernova distance ladder Hubble constant ($H_0$) measurement differs from the Planck prediction at 5$\sigma$ \citep{riess/etal:2022}. This discrepancy has not been convincingly resolved, either through identification of biases in the measurements, or modification to the cosmological model, despite intensive scrutiny over the past decade.

Various measurements of the growth of structure at late times are also in tension with Planck at up to the $2-3\sigma$ level, with the Planck \lcdm\ model predicting larger matter fluctuations than are inferred from the more direct observations. This is known as the `$S_8$' tension \citep[e.g.,][]{planck2013-p11,joudaki/etal:2016,hildebrandt/etal:2017,mccarthy/etal:2018,nguyen/huterer/wen:2023}. While this tension is not solely with weak lensing measurements, weak lensing analysis choices, including the treatment of intrinsic galaxy shape alignments, have been shown to moderately impact its significance \citep{des/kids:2023}.

\begin{table*}
  \caption{Posterior $A_L$ Constraints from Analyses of Planck Temperature and Polarization Data since 2018 Release}
  \label{table:existingAL}
  \centering
  \begin{tabular}{llllcc}
  \hline
  Reference&Data Version&Likelihood&Data Combination&$A_L$&`$N\sigma$' Preference\\&&&&&for $A_L>1$\\
  \hline
  \hline
  \cite{planck2016-l06}&PR3/2018&\texttt{plik}&TTTEEE+\texttt{lowl}/\texttt{lowE}&$1.180\pm0.065$&$2.8\sigma$\\
  &PR3/2018&\texttt{plik}&TT+\texttt{lowl}/\texttt{lowE}&$1.243\pm0.096$&$2.5\sigma$\\
  \hline
  \cite{rosenberg/gratton/efstathiou:2022}&PR3/2018&\texttt{CamSpec}&TTTEEE+\texttt{lowl}/\texttt{lowE}&$1.146\pm0.061$&$2.4\sigma$\\
  &PR3/2018&\texttt{CamSpec}&TT+\texttt{lowl}/\texttt{lowE}&$1.215\pm0.089$&$2.4\sigma$\\
  &PR4/NPIPE&\texttt{CamSpec}&TTTEEE+\texttt{lowl}/\texttt{lowE}&$1.095\pm0.056$&$1.7\sigma$\\
  &PR4/NPIPE&\texttt{CamSpec}&TT+\texttt{lowl}/\texttt{lowE}&$1.198\pm0.084$&$2.4\sigma$\\
  \hline
  \cite{tristram/etal:2024}&PR4/NPIPE&\texttt{HiLLiPoP}&TTTEEE+\texttt{lowl}/\texttt{LoLLiPoP}\footnote{\texttt{LoLLiPoP} is an alternative $\ell<30$ polarization likelihood \citep{tristram/etal:2021}.}&$1.036\pm0.051$&$0.7\sigma$\\
  &PR4/NPIPE&\texttt{HiLLiPoP}&TT+\texttt{lowl}/\texttt{LoLLiPoP}&$1.068\pm0.081$&$0.8\sigma$\\
  \hline
  \end{tabular}
\end{table*}

The Hubble tension cannot be resolved through any systematic error or problem specific to the Planck data. Baryon acoustic oscillation (BAO) measurements exhibit significant tension with the distance ladder within \lcdm\ when combined either with CMB measurements (both Planck and others), or with primordial deuterium abundance constraints, independent of CMB anisotropy measurements altogether \citep[e.g.,][]{aubourg/etal:2015,addison/etal:2018,alam/etal:2021}. \cite{addison/etal:2018} found $H_0=68.30\pm0.72$~km~s$^{-1}$~Mpc$^{-1}$ from combining Wilkinson Microwave Anisotropy Probe \citep[WMAP;][]{bennett/etal:2013} CMB measurements with BAO, for example. The Planck data do, however, play a central role in limiting or ruling out many models proposed to explain the discrepancy, particularly those that modify physics around recombination \citep[e.g.,][]{knox/millea:2020,smith/etal:2022}. More broadly, the Planck data will remain valuable, and carry a lot of statistical weight, for a host of future cosmological analyses. If residual systematic errors or problems do exist in the Planck data, identifying these issues and understanding their impact is clearly important.

The overall message from the Planck Collaboration in the final 2018 papers is that \lcdm\ continues to successfully describe the Planck measurements. Since the first Planck cosmological data release, however, the Planck data have shown several closely related internal tensions with \lcdm\ at the $2-3\sigma$ level, which have been discussed extensively in the Planck papers and subsequent literature.

CMB photons are deflected by gravitational lensing from large-scale structure and this effect can be measured by arcminute resolution CMB experiments \citep[e.g.,][]{das/etal:2011,vanengelen/etal:2012}. For the range of multipoles probed by Planck ($\ell<2500$), convolution with the lensing kernel leads to a smoothing out of the acoustic structure in the temperature spectrum, slightly lowering the peaks and filling in the troughs \citep[e.g.,][]{hu:2000}. On smaller angular scales there is also an overall enhancement of power due to transfer from the lower multipoles \citep[Figure~6 of][]{lewis/challinor:2006}, however this is a minor effect at the Planck noise levels.

The Planck 2018 temperature and polarization data show a $2.8\sigma$ preference for the phenomenological lensing amplitude parameter, $A_L$ \citep{calabrese/etal:2008}, to exceed the physically expected value of unity \citep{planck2016-l06}. The marginalized constraint is $A_L=1.180\pm0.065$.\footnote[3]{The 1D $A_L$ posteriors are reasonably approximated as Gaussian and so we quote mean and standard deviation (SD) throughout this work. The standard deviation and 68\% credible interval are similar in all cases.} This behavior is not present in the reconstruction of the lensing deflection field from higher-point statistics of the Planck maps. The Planck 2018 lensing reconstruction constraint on an overall phenomenological rescaling amplitude, relative to the full Planck \lcdm\ best fit model, is\footnote[4]{An analysis using the more recent PR4/NPIPE processing of the Planck data yields $A_{\phi\phi}=1.004\pm0.024$ \citep{carron/mirmelstein/lewis:2022}.} $A_{\phi\phi}=1.011\pm0.028$ \citep{planck2016-l08}.

We show constraints on $A_L$ from several published analyses of Planck data in Table~\ref{table:existingAL}. Throughout this paper, the labels `\texttt{lowl}' and `\texttt{lowE}' refer to the Planck 2018 low-multipole temperature and E-mode likelihoods covering $2\leq\ell\leq29$. The $\ell\geq30$ Planck polarization (TE and EE) spectra do not show a meaningful preference for $A_L>1$, although their impact when combined with TT depends more strongly on the data version and analysis choices \citep[e.g., Section~3.10 of][]{planck2016-l05}.

Since the Planck 2018 release, tighter cosmological parameter constraints have been obtained from the Planck data using the \texttt{CamSpec} likelihood \citep{efstathiou/gratton:2021,rosenberg/gratton/efstathiou:2022}. These analyses clean the majority of the dust foreground from the Planck 143 and 217~GHz maps using higher frequency data, and utilize larger sky fractions than the Planck 2018 analysis. The published 2018 likelihood, \texttt{plik}, instead used a range of foreground templates with amplitudes as nuisance parameters in the cosmological fitting. Compared to the PR3/2018 release, the PR4/NPIPE data reprocessing predominantly affected the Planck polarization data, but also led to a reduction in the temperature noise \citep{planck2020-LVII,rosenberg/gratton/efstathiou:2022}.

Recently, after the calculations for this work were completed, \cite{tristram/etal:2024} presented cosmological constraints from the PR4/NPIPE data using the \texttt{HiLLiPoP} and \texttt{LoLLiPoP} likelihoods. Their results are summarized in the bottom two rows of Table~\ref{table:existingAL} and feature $A_L$ constraints consistent with unity within $1\sigma$. Many analysis choices made by \cite{tristram/etal:2024} differ from those in  the Planck 2018 and \texttt{CamSpec} approaches (e.g., sky masks, foreground modeling, power spectrum covariance approximation), and consequently it is unclear what exactly is driving the difference in $A_L$ results from the earlier analyses of the Planck data.

Data from other high-resolution CMB experiments, specifically the Atacama Cosmology Telescope \citep[ACT; see][and references therein]{aiola/etal:2020} and the South Pole Telescope \cite[SPT; e.g.,][]{dutcher/etal:2021}, do not show a preference for $A_L>1$. The SPT data across three generations of receiver actually prefer $A_L<1$ at low significance \citep[up to $1.4\sigma$;][]{story/etal:2013,henning/etal:2018,dutcher/etal:2021}. The recent ACT Data Release 6 (DR6) lensing reconstruction has comparable precision to, and agrees well with, the Planck lensing reconstruction, with no evidence for a lensing amplitude excess over the \lcdm\ expectation \citep{madhavacheril/etal:2024}.

Within \lcdm, with $A_L=1$, the Planck temperature data at low and high multipoles prefer different values of some parameters, including $H_0$, and the spectrum amplitude $A_se^{-2\tau}$, at up to $2.9\sigma$ significance \citep[e.g.,][]{planck2014-a13,addison/etal:2016,planck2016-LI}. The difference across the full parameter space is not statistically significant. Here, `low multipoles' refers to $\ell<800$ or $\ell<1000$, roughly the range already measured by WMAP. Lensing primarily affects the higher multipoles, in the damping tail, and raising $A_L$ helps reconcile parameters between the low and high multipole ranges.

Another related internal Planck tension is the $\sim3\sigma$ preference for a closed universe ($\Omega_k<0$) when using Planck temperature and polarization data alone \citep[e.g.,][]{planck2016-l06,divalentino/melchiorri/silk:2020,handley:2021}. This resolves the multipole-dependent \lcdm\ parameter differences in a similar way to allowing $A_L>1$. Changing $\Omega_k$ changes the angular diameter distance to last scattering. For the primary CMB anisotropies, sourced at or around last scattering, this change can be effectively canceled with shifts in \lcdm\ parameters such as  $H_0$ and the matter density, $\Omega_m$, maintaining a good match to the acoustic structure measured at high precision in the power spectra. Changing $\Omega_k$ also impacts the post-recombination growth of structure, however, and the temperature and polarization spectra are primarily sensitive to this through the gravitational lensing effect. This allows $\Omega_k$ to mimic the effect of $A_L$ in fits that do not include more direct low-redshift cosmological measurements. While allowing $\Omega_k$ to vary improves the fit to the Planck temperature and polarization data, it should be emphasized that the resulting values of other parameters are discrepant with many other data sets. For example\footnote[5]{Full cosmological parameter tables from the 2018 data release are available on the Planck Legacy Archive \url{http://pla.esac.esa.int/pla/\#cosmology}.}, the Hubble constant is constrained to be $H_0=54.4^{+3.3}_{-4.0}$~km~s$^{-1}$~Mpc$^{-1}$. While $\Omega_k$ is a physical model parameter, unlike the phenomenological $A_L$, the solution favored by the Planck 2018 temperature and polarization data alone with $\Omega_k$ varying is not a convincing description of the universe, given other measurements.

The preference for $A_L>1$, the preference for different \lcdm\ parameter values from different multipole ranges, and the preference for $\Omega_k<0$ in the Planck temperature and polarization data are not independent issues but better viewed as three symptoms of some common underlying effect. Given the lack of lensing amplitude anomaly in the lensing reconstruction, or in any other CMB dataset, the preference for $A_L>1$ in the Planck TT data should not be interpreted as a preference for actual additional physical lensing (and similarly for $\Omega_k$, as mentioned above). This point has been made consistently by the Planck Collaboration \citep[e.g.,][]{planck2016-l05}, but is worth reiterating. We instead view $A_L$ as a proxy for some other (unknown) phenomenon or phenomena, although whether the main cause is cosmological, systematic, or just a statistical fluke, has not been clear in the past.

In this paper, we attempt to shed light on the origin of the $A_L$ anomaly by first reproducing the Planck 2018 TT results, and then performing additional tests of the temperature data, specifically focusing on the choice of sky mask and frequency channels. We use PR3/2018 Planck products\footnote[6]{Available on the Planck Legacy Archive, \url{https://pla.esac.esa.int/\#maps}.}, which have been discussed and utilized extensively within the cosmological community.

In Section~\ref{sec:computations}, we outline the steps we take to obtain cosmological constraints from the Planck temperature maps and present our reproduction of the Planck 2018 $\lcdm+A_L$ constraints. In Section~\ref{sec:ecliptic}, we present results from additional masking based on ecliptic latitude. In Section~\ref{sec:dust}, we present results from extending the Planck masks using the 857~GHz map to remove some additional regions with bright dust emission. In Section~\ref{sec:lowl}, we examine the impact of the $\ell<30$ TT multipoles on $A_L$. In Section~\ref{sec:additional}, we check the possible impact of residual CO emission or differences in effective beam window function with sky cut. In Section~\ref{sec:Ok} we shows results varying $\Omega_k$ instead of $A_L$. A summary and conclusions follow in Section~\ref{sec:conclusions}.

Given the large number of different tests performed in this work, we provide a numbered list of main results, with pointers back to the relevant sections and subsections, at the start of Section~\ref{sec:conclusions}.

\section{Computing Cosmological Parameters from Planck Temperature Maps}
\label{sec:computations}

In this Section we describe our approach for calculating angular power spectra and power spectrum covariance matrices from the Planck 2018 data products, and modifying the public likelihood files for cosmological parameter fitting. We generally follow the Planck 2018 analysis \citep[hereafter PL18]{planck2016-l05}, however there are differences, in part because the Planck analysis cannot be perfectly reproduced from the public products. We then demonstrate that despite these differences we reproduce the Planck 2018 TT+\texttt{lowl}/\texttt{lowE} cosmological parameter constraints in the \lcdm+$A_L$ model to $0.1\sigma$ or better. We do not perform a reanalysis of the $\ell<30$ data or modify the \texttt{lowl}/\texttt{lowE} likelihoods.

In later Sections we modify the masks applied to the Planck maps. In each of those cases we repeat the steps described in this Section, in particular re-computing the mode-coupling matrices and all dependent quantities for each mask.

\cite{li/etal:2023} also reproduced the Planck 2018 parameter results to comparable precision, despite making slightly different assumptions (e.g., for the noise modeling). \cite{li/etal:2023} included polarization data, whereas our analysis is restricted to temperature. Given the complexity of the Planck data it is encouraging that multiple groups have been able to successfully recover the constraints published by the Planck Collaboration from the public data products.

\subsection{Maps and Masks}
\label{sec:maps_masks}

We follow PL18 in using the 100, 143, and 217~GHz half-mission (HM) maps to compute angular power spectra for use in cosmological parameter fitting, and use the 143$\times$217~GHz cross-spectra in addition to the three same-frequency spectra. We similarly only use HM1$\times$HM2 cross-spectra to avoid auto-spectrum noise bias and co-temporal systematic effects. For the 143$\times$217~GHz cross-spectrum we simply take the mean of the 143HM1$\times$217HM2 and 143HM2$\times$217HM1 spectra, since we find the improvement from the weighted linear combination used by PL18 is negligible.

The Planck 2018 temperature masks were constructed in the same way as in the 2015 analysis and are described in Appendix~A of \cite{planck2014-a13}. The final masks are constructed by combining (multiplying) a number of separate masks: (1) Galactic mask based on a coarsely-smoothed thresholding on 353~GHz intensity after a CMB subtraction, (2) point source mask constructed separately at each frequency from catalogs of bright point-like sources (both Galactic and extragalactic), (3) extended source mask targeting a small number of objects such as the Magellanic clouds and nearby spatially extended spiral galaxies, (4) missing pixel mask assigning zero weight to pixels that were not observed in a given frequency and HM map, and (5) CO mask removing bright Galactic CO line emission from the $J=1\to0$ and $J=2\to1$ transitions around 115 and 230~GHz. The CO mask is applied at only 100 and 217~GHz, since the 143~GHz bandpass does not overlap any CO lines. With the exception of the missing pixel masks, the masks are apodized to reduce mode-coupling from large scales to small scales. The Galactic masks depend on frequency, with around 40\% of the sky being removed (before apodization) at 217~GHz, down to around 20\% at 100~GHz, where the dust signal is fainter.

We made one addition to the public masks. We found a very small number of pixels (order tens) in the HM2 maps with reported variances that are either negative, or many orders of magnitude larger than the other pixels. Inspection of the corresponding entries in the data maps showed no obvious problems, but, if ignored, these anomalous variance entries disproportionately impact estimation of the noise contribution to the power spectrum covariance. We therefore assign these pixels zero weight, effectively treating them like extra missing pixels.

The decision to use a separate mask for each frequency and HM in the Planck 2018 analysis carries a computational cost because it requires computing many more matrices for building the various blocks of the power spectrum covariance (Section~\ref{sec:cov}, below). This cost is, however, negligible compared to computing requirements for the full Planck data analysis pipeline (e.g., the map making). We investigated using a single mask per frequency, combining the missing pixels across both HM masks. Unfortunately, while the missing pixels are a small fraction of the total pixels in the maps (at most around 0.2\% for 217~GHz HM2), the extra small-scale power in the combined masks does moderately impact power spectrum estimation at the higher multipole range ($\ell>1500$) due to increased coupling of power from large scales ($\ell<100$). We therefore follow PL18 and use the separate HM masks.

\subsection{Pseudo-Spectra and Mask Deconvolution}
\label{sec:pseudoCl}

We compute the so-called pseudo-spectra \citep[uncorrected angular power spectra of the data maps multiplied by the masks; e.g.,][]{hivon/etal:2002} using the \texttt{anafast} routine of the \texttt{healpy} Python implementation of the Hierarchical Equal Area isoLatitude Pixelization library\footnote[7]{\url{https://healpix.sourceforge.io/}} \citep[\texttt{HEALPix};][]{gorski2005}. Following the Planck analyses we do not apply additional weighting based on noise variance. We remove a zero-point (monopole) from each map, estimated as $\sum_pm_pw_p/\sum_pw_p$, where $m$ and $w$ are the maps and masks, and $p$ labels the pixels \citep{efstathiou/gratton:2021}. We investigated removing the dipole modes in a similar way but found this had no impact on the cosmological results. Note that the Planck 2018 maps are provided with the CMB dipole already removed. 

We correct the pseudo-spectra for the effect of the mask using the standard mode-coupling matrix \citep{hivon/etal:2002}, which depends on the angular cross-spectrum of the two masks for each frequency/HM pairing \citep[Appendix~C.1.2 of ][]{planck2014-a11}. We verified that our computation of the mode-coupling matrices matches that from the public \texttt{NaMaster}\footnote[8]{\url{https://github.com/LSSTDESC/NaMaster}} code \citep{alonso/etal:2019}. While the maximum multipole of the deconvolved spectrum used in the cosmological fitting is $\ell_{\rm max}=2508$ we compute spectra and matrices out to higher multipoles, typically $\ell_{\rm max}=3508$.

The deconvolved spectra are finally corrected for the beam and pixel window functions, using the public Planck 1D effective beam window function for each frequency/HM pairing. For the cosmological likelihood we used binned spectra, following the Planck binning scheme, which compresses multipoles over the range $30\leq\ell\leq2508$ into 215 bandpowers with widths varying from $\Delta\ell=5$ to 33, with a restricted multipole range at 100 and 143~GHz that discards noisy data at high multipoles (Equation~22 and Table~12 of PL18).

\subsection{Power Spectrum Covariance Matrices}
\label{sec:cov}

Accurate modeling of the power spectrum covariance matrices is clearly important for robust cosmological parameter inference. The task is essentially to extend the simple expression valid for the auto-power-spectrum in the full-sky, white noise case,
\be
\left\langle\Delta \hat{C}_{\ell}\Delta \hat{C}_{\ell'}\right\rangle=\frac{2\delta_{\ell\ell'}}{2\ell+1}\left(C_{\ell}+N_{\ell}/b_{\ell}^2\right)^2,
\ee
where $N_{\ell}$ is the noise power and $b_{\ell}$ is the beam profile. The final expression must account for the fact that we are constructing power spectra from the cross-correlation of HM1 and HM2 map pairs, and for the masking and anisotropic noise, which impact the variance at each $\ell$ as well as introducing off-diagonal contributions. Full details may be found in Appendix~C.1. of \cite{planck2014-a11}, and references therein.

We handle two aspects of the covariance calculation differently from the Planck Collaboration and discuss these below.

\subsubsection{Improved Narrow-Kernel Approximation}
\label{sec:NKA}

In the Planck analyses, the covariance is estimated using the so-called Narrow-Kernel Approximation \citep[e.g.,][]{efstathiou:2004,nicola/etal:2021}. This enormously simplifies the full set of summations over $\ell$ and $m$ indices but is inaccurate at around the 10\% level for some multipoles, due to the combination of the very red nature of the CMB power spectrum\footnote[9]{Recall that the power on the sky is $C_{\ell}$, not the conventionally plotted quantity $\ell(\ell+1)C_{\ell}$, and decreases by many orders of magnitude over the multipole range of interest.}, and the presence of small-scale structures in the masks \citep{planck2014-a11,huang/etal:2018}.

The Planck Collaboration used simulations to correct for this inaccuracy \citep[Appendix~C.1.4. of][]{planck2014-a11}. A separate correction is needed for each mask. Given the large number of masks we use in this work, we considered alternative approaches and found the Improved Narrow-Kernel Approximation \citep{nicola/etal:2021}, originally proposed by \cite{gruetjen/etal:2017}, provided the required accuracy with minimal additional computing cost. Each occurrence of the fiducial sky spectrum in the Narrow-Kernel covariance approximation is replaced with the fiducial pseudo-spectrum divided by the sky fraction of the appropriate mask product. That is, 
\be
\left.C^{AB}_{\ell}\to\sum_{\ell'}M^{AB}_{\ell\ell'}C^{AB}_{\ell'}\middle/\frac{1}{N_{\rm pix}}\sum_pw^A_pw^B_p\right.,
\ee
where $A$ and $B$ label the pair of maps, $M_{\ell\ell'}$ is the mode-coupling matrix (Section~\ref{sec:pseudoCl}), and $w^A$ and $w^B$ are the masks.

For the fiducial sky model required in the covariance calculation we used the best-fit \lcdm\ CMB-plus-foreground model from the public Planck 2018 TTTEEE+\texttt{lowl}/\texttt{lowE} chains. We emphasize that this choice does not have a significant impact on cosmological parameters, and that other models that provide a reasonable fit to the data yield effectively indistinguishable results. For example, using the best-fit TT+\texttt{lowl}/\texttt{lowE} \lcdm\ model, or best-fit TT+\texttt{lowl}/\texttt{lowE} \lcdm$+A_L$ model, results in negligible shifts in cosmological parameters.

While the $100\times143$ and $100\times217$~GHz spectra are not used in the cosmological fitting, we require a fiducial model for these spectra for some blocks of the covariance matrix. We adopted a simple empirical approach. We first assumed 100\% coherence between the foregrounds at 100~GHz and the higher frequencies. We then compared to the $100\times143$ and $100\times217$~GHz cross-spectra measured using the real data and rescaled the $100\times217$~GHz spectra by a factor $1-0.03(\ell/1000)^2$ to better match the measurements at higher multipoles. A difference of this sort is not unexpected because the point source contribution violates the assumption of 100\% foreground coherence, being dominated by synchrotron sources at 100~GHz but with more dust emission at 217~GHz \citep[e.g.,][]{reichardt/etal:2012}. This correction, and the exact treatment of the 100~GHz cross-spectra, have negligible impact on cosmological parameters.

\subsubsection{Noise Modeling}
\label{sec:noise}

The Planck Collaboration used empirical parametric models to describe the deviations from white noise power as a function of multipole for the pseudo-spectrum covariance matrices (Equation~C.24 and surrounding text in PL18). The final noise model is a combination of (1) a diagonal pixel-space variance (the white noise levels estimated during map making and provided with the public Planck maps), and (2) an isotropic ($m$-independent) harmonic-space rescaling to capture the multipole-dependent structure of the real noise.

We follow this general approach, but use a different method to estimate the harmonic-space rescaling factors for each frequency and HM, which we denote $\psi_{\ell}^{\nu,\textrm{HM}}$. PL18 used further splits of the HM maps into half-ring difference maps (so effectively quarter-mission splits). While full-mission half-ring maps are publicly available, the HM half-ring maps used for this noise modeling step are not. Additionally, the HM half-ring maps were found to contain correlated noise that required an extra correction (Section~3.3.3 of PL18). Because of these issues we model the deviations from white noise more directly using only the HM data maps and HM Full-Focal Plane (FFP10) simulations provided by the Planck Collaboration \citep[see][PL18]{planck2014-a14}.

We first compute noise power spectra from the FFP10 noise-only simulations for each frequency and HM, for each of the 300 realizations provided. We use the same masks as in the data analysis (Section~\ref{sec:maps_masks}). We estimate $\psi_{\ell}^{\nu,\textrm{HM}}$ from dividing the mean of the FFP10 noise pseudo-spectra by the white noise pseudo-spectra expected given the pixel variance maps and masks.

This approach for computing $\psi_{\ell}^{\nu,\textrm{HM}}$, using the FFP10 noise simulations alone, yields cosmological parameter results in excellent agreement with those we present in this paper, with differences around $0.1\sigma$ or less, far smaller than the $A_L$ effect we are investigating. However, the FFP10 noise simulations are known not to perfectly capture the noise properties of the real data \citep{planck2014-a14}, with discrepancies up to around 5\% in units of power for the TT HM spectra on scales where the noise contributes significantly to the error bars. We therefore perform a leading-order correction for this mismatch using a low-order polynomial rescaling of $\psi_{\ell}^{\nu}$ to better match the data noise properties. More details on this correction are provided in Appendix~\ref{app:noise}.

\subsection{Cosmological Parameter Fitting and Modified Likelihood Files}

We use the \texttt{CAMB}\footnote[10]{\url{https://camb.info/}} code \citep{lewis/challinor/lasenby:2000,howlett/etal:2012} to compute theory power spectra from cosmological parameters, and \texttt{CosmoMC}\footnote[11]{\url{https://cosmologist.info/cosmomc/}} \citep{lewis/bridle:2002} to read our modified likelihood files and perform MCMC sampling. For reproducing the Planck results we use the same priors and convergence settings as in PL18. We modify some priors when changing the masks in the later sections (notably removing the Gaussian prior on the 217~GHz Galactic dust amplitude, since this is specific to the Planck mask).

\begin{figure*}
\centering
\includegraphics[]{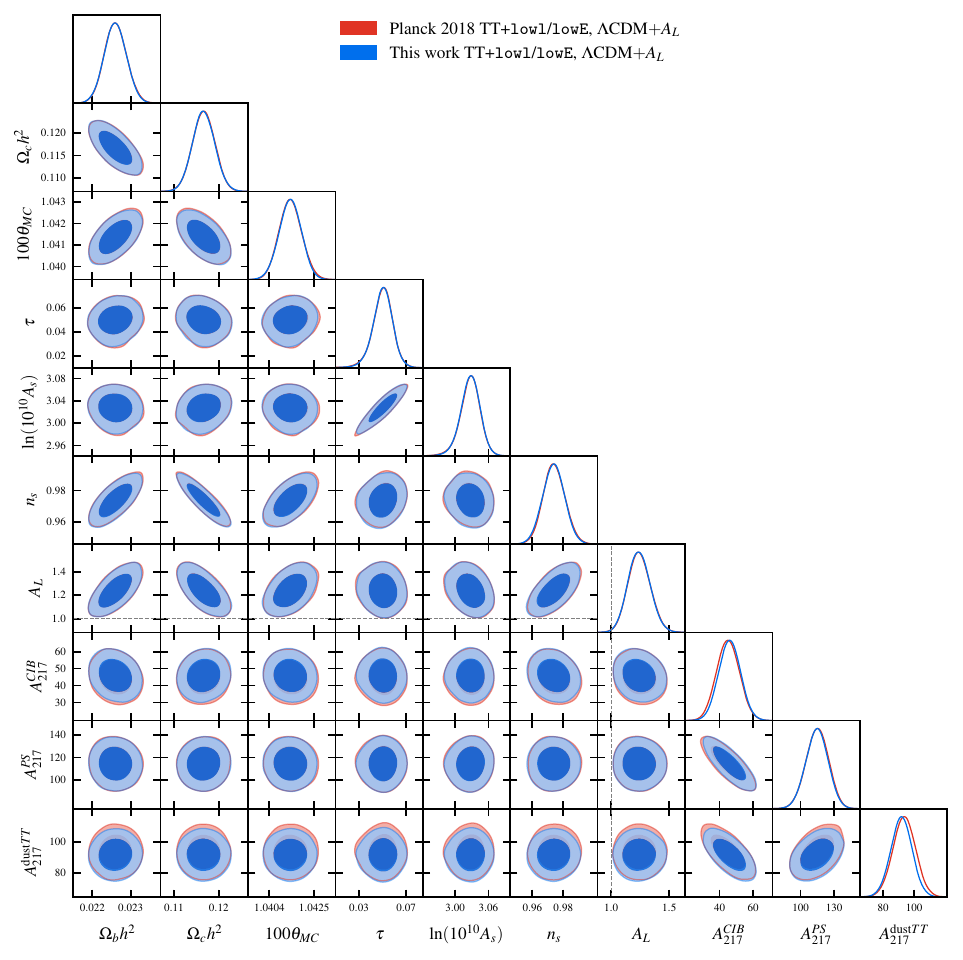}
\caption{Reproduction of the Planck 2018 TT+\texttt{lowl}/\texttt{lowE} constraints in the \lcdm+$A_L$ model. Despite small differences in the masking (Section~\ref{sec:maps_masks}), covariance approximation (\ref{sec:NKA}), and noise modeling (\ref{sec:noise}), our likelihood matches the Planck results to $0.1\sigma$ or better for cosmological parameters. We also show the main 217~GHz foreground parameters, where shifts are larger,  up to $0.25\sigma$ for the 217~GHz Galactic dust amplitude ($A_{217}^{\textrm{dust}TT}$), though still subdominant to statistical uncertainty. Agreement for other foreground parameters, for fits to individual TT spectra, and for the \lcdm\ model, is at least as good as that shown here.}
\label{fig:triangle_reproduce}
\end{figure*}

Each Planck 2018 \texttt{plik} likelihood instance consists of a set of files contained in a folder with the .clik extension. There is one likelihood for TT, one for TT without multipole binning, one for TTTEEE, and so on. The \texttt{CosmoMC} code reads the contents of the specified folder at run-time (in addition to other requested likelihoods, for example the \texttt{lowl} and \texttt{lowE} likelihoods). Our philosophy is to make minimal edits to the Planck files, and no edits at all to the code used for reading the files, to reduce the possibility of new errors or issues being introduced. To perform MCMC fits with modified likelihoods, we make a copy of one of the original .clik folders and edit the relevant files. In this work we only ever edit the power spectrum vector, containing the binned spectra, and covariance matrix. In the .clik folders it is actually the inverse covariance matrix that is stored.\footnote[12]{Note that the covariance matrices are stored in a different format in the foreground-marginalized \texttt{pliklite} likelihoods. Since we are interested in changing the masking and other analysis choices we do not use the \texttt{pliklite} likelihood in this work.} Cases where we remove one or more of the original spectra ($100\times100$, $143\times143$, $143\times217$, and $217\times217$~GHz) are straightforward to handle. For example, when fitting the $217\times217$~GHz spectrum alone, we replace the $217\times217$~GHz block of the inverse covariance matrix with the inverse of our $217\times217$~GHz matrix, and set the remaining elements to zero.

\subsection{Reproducing Planck 2018 Results}

Figure~\ref{fig:triangle_reproduce} compares \lcdm+$A_L$ parameter posteriors from the original Planck 2018 chains and from fitting our version of the $\ell\geq30$ TT likelihood. The Planck 2018 \texttt{lowl}/\texttt{lowE} likelihoods are included in both cases. Cosmological parameters agree to $0.1\sigma$ or better\footnote[13]{Since these parameters were obtained from essentially the same data we quote shifts in the posterior mean relative to the original Planck 2018 error bars rather than attempting to fold in both sets of uncertainties.} and 1D and 2D posteriors are in similarly good agreement. This indicates that small differences between our likelihood and the official Planck 2018 likelihood do not meaningfully impact cosmological parameter recovery. 

There are larger shifts for several foreground parameters, up to $0.25\sigma$ for the Galactic dust amplitude parameter at 217~GHz\footnote[14]{Note that a Gaussian prior of $91.9\pm20.0$~$\mu$K$^2$ is applied on this parameter in both fits (Section~3.3.2 of PL18). We verified that removing this prior does not significantly impact the parameter comparison.}, $A_{217}^{\rm dustTT}$. We show this and the other main 217~GHz foreground parameters, the amplitudes of the cosmic infrared background ($A^{CIB}_{217}$) and point source ($A^{PS}_{217}$) terms, in Figure~\ref{fig:triangle_reproduce}. Shifts are smaller for the foreground parameters associated with other frequencies. In general we find that the foreground parameters are slightly more sensitive to choices like the covariance approximations and noise modeling, while the cosmological parameters are more robust. Additionally, when we fit only 217~GHz, or the other spectra separately, the consistency with the corresponding \texttt{plik} constraints improves, indicating that the shift in $A_{217}^{\rm dustTT}$ in Figure~\ref{fig:triangle_reproduce} is not a symptom of a meaningful mismatch in the actual 217~GHz spectra.

Our power spectra at 100, 143 and 217~GHz agree well with the \texttt{plik} spectra, with deviations of at most $0.04$ times the quoted \texttt{plik} uncertainty in any given bin. At $143\times217$~GHz the differences are larger, up to around $0.4\sigma$ at small scales. This is due to our simply taking the mean of the 143HM1$\times$217HM2 and 143HM2$\times$217HM1 spectra, as opposed to the optimally noise-weighted combination used in \texttt{plik}. This has negligible impact on parameters, and our $143\times217$~GHz cosmological and foreground parameter constraints agree with those from the $143\times217$~GHz \texttt{plik} spectrum as well or better than for the multifrequency case shown in Figure~\ref{fig:triangle_reproduce}.

The best-fit $\chi^2$ from our $\ell\geq30$ TT likelihood in the \lcdm+$A_L$ model is $728.5$, compared to $752.9$ with \texttt{plik}.\footnote[15]{The best-fit parameters and $\chi^2$ are obtained from a dedicated maximization of the posterior probability using the \texttt{action=2} setting in \texttt{CosmoMC}.} Neither $\chi^2$ value is anomalous given the number of degrees of freedom, which we estimate at 752.\footnote[16]{There are 765 power spectrum bins, 7 cosmological parameters and 15 foreground/nuisance parameters. Of the latter, 8 are either unconstrained, or constrained primarily by priors. Additionally, $\tau$ is primarily constrained by the \texttt{lowE} likelihood. Estimating the number of degrees of freedom as 752, the probability to exceed (PTE) for our likelihood is 0.72, and for the \texttt{plik} likelihood is 0.48.} The difference is similar in \lcdm. The $\chi^2$ values are sensitive to differences in the likelihood approximations that negligibly impact parameter constraints (e.g., the exact details of the noise modeling). We are primarily concerned with the parameters, and have not investigated the difference in $\chi^2$ with respect to \texttt{plik} in more detail. However, we did perform simulations to quantify possible biases in our $\chi^2$ from approximations described earlier in this Section. These simulations consist of 500 realizations of CMB, noise and foreground fluctuations for each of the 100, 143 and 217~GHz channels and each HM. We find no measurable level of bias in the simulation $\chi^2$ distribution, and estimate the possible bias in the data $\chi^2$ due to our covariance matrix approximations at $\Delta\chi^2\lesssim5$, several times smaller than the difference we see with \texttt{plik}. More details of these simulations are provided in Appendix~\ref{app:sims}.

\subsection{Frequency Dependence of Planck 2018 $A_L$ Temperature Results}

Before turning to masking changes, we discuss an important aspect of the $A_L$ constraints from the Planck 2018 temperature data: frequency dependence. PL18 claim that the $A_L$ deviation appears coherent across frequencies, based on results from removing one TT spectrum or frequency at a time (see their Section~3.10 and Figure~83). This would seemingly rule out many potential systematic origins, at least as the principal explanation for the $A_L$ anomaly.

To examine possible frequency dependence, we fit $\lcdm+A_L$ to each of the four TT spectra in the \texttt{plik} likelihood separately. Table~\ref{table:plikAL} shows $A_L$ constraints from these fits. Note that for consistency with later values we show results using our likelihoods here, but that using the Planck 2018 likelihoods would make no meaningful difference. The 217~GHz spectrum alone prefers $A_L>1$ at around $2.4\sigma$, comparable to the significance in the combined case. The other spectra are consistent with $A_L=1$ within $2\sigma$. Additionally, the $A_L$ posterior mean for 217~GHz is notably higher than the other frequencies, and the combined fit. A similar constraint, centered at around 1.35, is shown in Figure~83 of PL18 for the case when the 143~GHz data are removed.

In the single-spectrum fits the foreground parameters are less well constrained and more strongly correlated with the cosmological parameters. To check the impact of this on $A_L$, for example, whether some kind of foreground parameter volume effect is driving the frequency dependence of the single-spectrum fits, we performed fits with the foreground and nuisance parameters fixed. For these fits we used the best-fit foreground values from the \lcdm\ fits to all the TT spectra, with $A_L$ fixed to 1 (i.e., the baseline Planck TT+\texttt{lowl}/\texttt{lowE} values). These results are shown in the third column of Table~\ref{table:plikAL}. While the preferred $A_L$ values do shift downwards, the frequency dependence of the single-spectrum results does not disappear.

In light of these results, our overall interpretation regarding frequency dependence of $A_L$ differs from that of PL18: Table~\ref{table:plikAL} shows that the $A_L$ anomaly does have some amount of frequency dependence. In PL18, the argument is that the shifts in $A_L$ from removing one TT spectrum or frequency are not statistically significant. In our view, however, even some low-level frequency dependence seems worth examining, given we are investigating an effect that is only $2.5\sigma$ in the TT data to begin with. In the Sections below we therefore show and discuss results for the individual spectra as well as the combined case.

\begin{table}
\caption{Posterior $A_L$ Constraints from Fits to our Planck TT Likelihoods plus the \texttt{lowl}/\texttt{lowE} Likelihoods}
\begin{tabular}{lcc}
\hline
TT Spectrum&$A_L$&$A_L$, Fixed FG\footnote{Fixing the \texttt{plik} foreground (FG) parameters to the best-fit values from the \lcdm\ fit to all the spectra}\\
\hline
\hline
100~GHz&$1.171\pm0.180$&$1.098\pm0.159$\\
143~GHz&$1.171\pm0.113$&$1.121\pm0.084$\\
$143\times217$~GHz&$1.194\pm0.112$&$1.156\pm0.082$\\
217~GHz&$1.362\pm0.147$&$1.283\pm0.098$\\
All&$1.240\pm0.094$&$1.167\pm0.073$\\
\hline
\end{tabular}
\label{table:plikAL}
\end{table}

\section{Additional Masking based on Ecliptic Latitude}
\label{sec:ecliptic}

In this Section we investigate how $A_L$ from the Planck 2018 temperature maps is impacted by additional masking based on ecliptic latitude. The Planck scan strategy involved deepest observations of regions near the ecliptic poles, which were scanned at a wide range of orientations \citep{planck2011-1.1}. Pixels near the ecliptic had fewest observations, and were scanned at a narrow range of orientations roughly perpendicular to the ecliptic. Additionally, various Solar System objects contribute potential microwave foreground signals along or near the ecliptic (e.g., planets, zodiacal light from interplanetary dust grains). Examining parameter consistency from splitting the sky by ecliptic latitude therefore tests the treatment of various effects in the Planck data reduction and modeling. An internal consistency test based on ecliptic cuts was described in the Planck 2018 lensing reconstruction analysis \cite[Section~4.2 of][]{planck2016-l08}.

We constructed masks that remove pixels by ecliptic latitude, either masking outwards around the poles, or outwards symmetrically about the equator. In each case the ecliptic masks were apodized with a $\sigma=2\degree$ Gaussian, following the procedure described in Appendix~\ref{app:noise} of \cite{planck2014-a11} to ensure pixels with zero weight are not assigned some non-zero value in the smoothing process. We chose cuts of $40\degree$ and $60\degree$ for masking around the poles, and $15\degree$ and $30\degree$ for masking around the ecliptic equator. These correspond to removal of roughly a quarter or roughly half of the sky. We denote these masks \texttt{pole40}, \texttt{pole60}, \texttt{equa15} and \texttt{equa30}. These ecliptic masks were then multiplied by the frequency dependent masks described in Section~\ref{sec:maps_masks}. For each ecliptic cut and new set of masks, all the HM spectra and covariance matrices were recomputed following the approach described in the previous Section. 

For the MCMC results shown below we adopted the same priors as in the reproduction of the Planck 2018 results, except that we remove the Gaussian prior on the 217~GHz Galactic dust amplitude, $A_{217}^{\textrm{dust}TT}$, which was computed specifically for the \texttt{plik} 217~GHz mask from $217\times545$~GHz cross-correlations (Section~3.3.2 of PL18). We find that the 217~GHz power changes significantly with the ecliptic masking, and so allow $A_{217}^{\textrm{dust}TT}$ to vary freely. This is discussed further in Section~\ref{sec:ecl_power}, below.

Separately, we investigated modifying the shape (multipole dependence) of the 217~GHz Galactic dust power spectrum templates for the different masks, largely following the procedure described by PL18. We found that this negligibly impacted $A_L$, except for the most extreme dust masks discussed in Section~4 below, where the effect was up to a $0.2-0.3\sigma$ shift. For simplicity we therefore report results with the original dust templates. We also tested the effect of updating the 217~GHz foreground model assumed for the power spectrum covariance matrix as the mask was changed. This also had a negligible effect on $A_L$. For more details of these tests see Appendix~\ref{app:dusttemplates}.

\begin{figure*}
\centering
\includegraphics[]{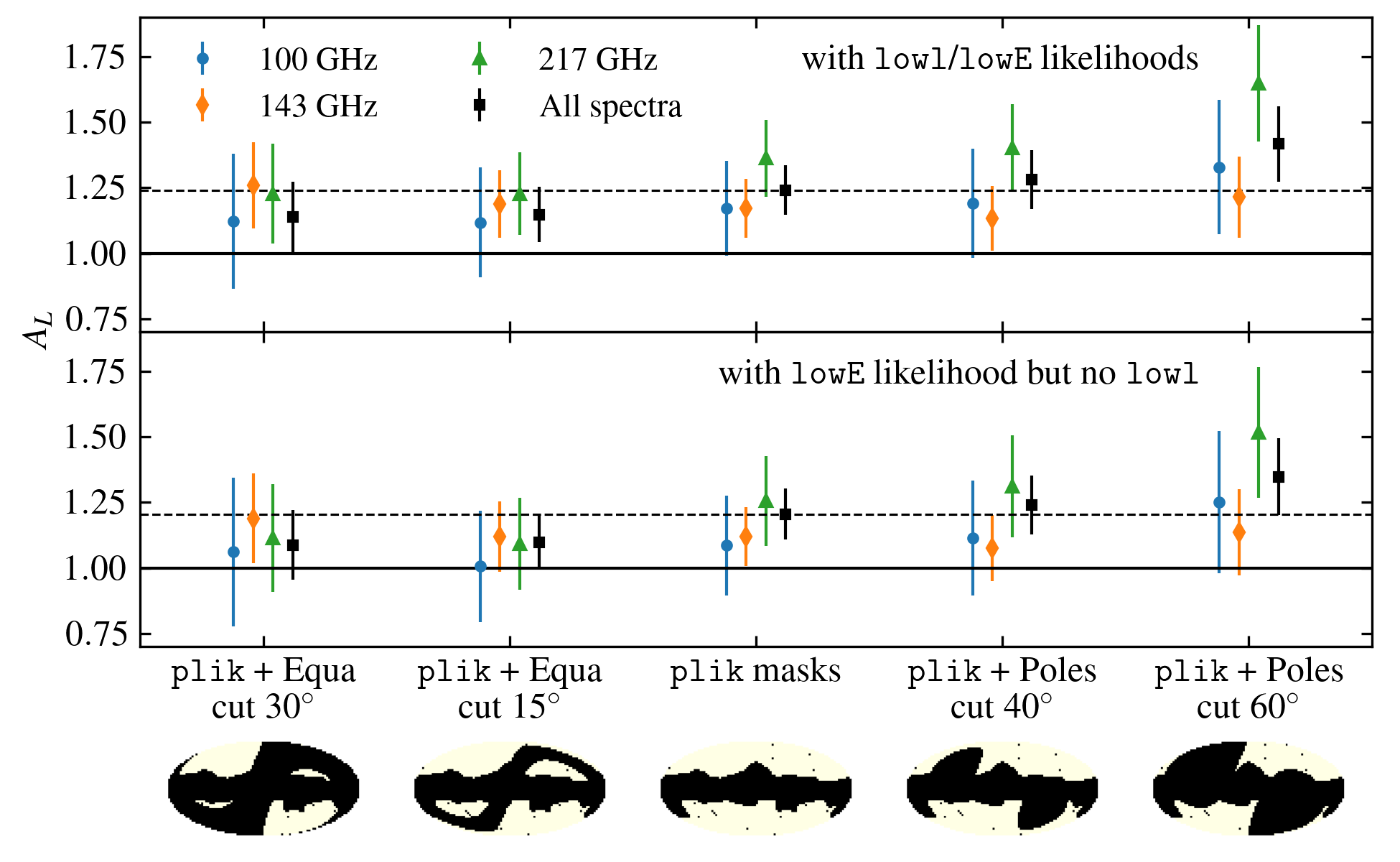}
\caption{The preference for $A_L>1$ in the Planck temperature data is strongest at low ecliptic latitudes, particularly for 217~GHz. We show marginalized $A_L$ constraints for the original Planck 2018 \texttt{plik} masks, and with additional masking around the ecliptic and ecliptic poles, indicated visually for the 217~GHz mask in the Mollweide images. (\textsl{Top:}) Results including the \texttt{lowl}/\texttt{lowE} 2018 likelihoods covering multipoles $\ell<30$. (\textsl{Bottom:}) Results without the $\ell<30$ TT likelihood, \texttt{lowl}, to highlight the contribution these low multipoles make to the preference for $A_L>1$. Points are offset horizontally for each mask for visual clarity. Horizontal dashed lines indicate the mean $A_L$ values for the fits using \texttt{plik} masks with all frequencies, 1.240 and 1.205 for the top and bottom panels, respectively.}
\label{fig:ecliptic_AL_1d}
\end{figure*}

\subsection{$A_L$ Constraints with Ecliptic Masks}
\label{sec:ALecliptic}

Figure~\ref{fig:ecliptic_AL_1d} shows 1D marginalized $A_L$ posterior constraints from varying the ecliptic masking. We show results for each frequency separately, as well as for the combined multifrequency (`All spectra') fit, which also includes the $143\times217$~GHz cross-spectrum. The left-most set of points use only regions near the ecliptic poles, while at the other extreme the right-most points use only the sky at low ecliptic latitudes. The Mollweide images show the unapodized 217~GHz masks. Note that the 100 and 143~GHz masks involve less masking around the Galactic plane (Section~\ref{sec:noise}). The main result from Figure~\ref{fig:ecliptic_AL_1d} is that the $A_L$ values from the 217~GHz and multifrequency fits increase from left to right (i.e., as the sky near the ecliptic is given more weight). The trend is weaker or absent in the lower frequencies. We report the $A_L$ posterior means and standard deviations, and sky fractions per mask, in Table~\ref{table:ecl_mask}.

\begin{table*}
  \centering
  \caption{Dependence of $A_L$ from Planck Temperature Data on Ecliptic Latitude Masks and $\ell<30$ \texttt{lowl} TT Likelihood}
  \begin{tabular}{llccc}
\hline
Mask&TT Spectrum&$f_{\rm sky}$&$A_L$&$A_L$\\
&&&(with \texttt{lowl}, \texttt{lowE})&(with \texttt{lowE} only)\\
\hline
\hline
\texttt{plik} + \texttt{equa30}&100~GHz&0.30&$1.121\pm0.258$&$1.060\pm0.283$\\
&143~GHz&0.26&$1.259\pm0.164$&$1.188\pm0.171$\\
&217~GHz&0.22&$1.227\pm0.191$&$1.113\pm0.205$\\
&All&various&$1.138\pm0.134$&$1.087\pm0.133$\\
\hline
\texttt{plik} + \texttt{equa15}&100~GHz&0.47&$1.117\pm0.209$&$1.006\pm0.212$\\
&143~GHz&0.41&$1.189\pm0.129$&$1.119\pm0.133$\\
&217~GHz&0.33&$1.226\pm0.157$&$1.092\pm0.176$\\
&All&various&$1.147\pm0.106$&$1.097\pm0.105$\\
\hline
\texttt{plik}&100~GHz&0.69&$1.171\pm0.180$&$1.085\pm0.190$\\
&143~GHz&0.60&$1.171\pm0.113$&$1.119\pm0.113$\\
&217~GHz&0.50&$1.362\pm0.147$&$1.256\pm0.172$\\
&All&various&$1.240\pm0.094$&$1.205\pm0.096$\\
\hline
\texttt{plik} + \texttt{pole40}&100~GHz&0.52&$1.190\pm0.208$&$1.114\pm0.218$\\
&143~GHz&0.45&$1.133\pm0.123$&$1.076\pm0.127$\\
&217~GHz&0.38&$1.402\pm0.167$&$1.312\pm0.194$\\
&All&various&$1.281\pm0.111$&$1.239\pm0.112$\\
\hline
\texttt{plik} + \texttt{pole60}&100~GHz&0.33&$1.328\pm0.256$&$1.250\pm0.272$\\
&143~GHz&0.29&$1.214\pm0.156$&$1.135\pm0.165$\\
&217~GHz&0.25&$1.647\pm0.222$&$1.516\pm0.250$\\
&All&various&$1.417\pm0.144$&$1.348\pm0.147$\\
\hline
\end{tabular}
\label{table:ecl_mask}
\end{table*}

The bottom panel of Figure~\ref{fig:ecliptic_AL_1d} shows results from removing the $\ell<30$ TT likelihood. This shifts $A_L$ downwards in all cases. Removing part of the data also increases parameter uncertainties, although this is a secondary effect. While the impact of lensing is negligible at $\ell<30$, increasing $A_L$ allows other parameters (e.g., $A_s$, $n_s$) to shift, improving the fit to the low multipoles while preserving a reasonable fit at higher multipoles in a way that is not possible in \lcdm. This indirect impact of the $\ell<30$ temperature data on $A_L$ has been discussed by the Planck Collaboration \citep[e.g.,][]{planck2014-a11,planck2016-LI}. We provide further discussion of this effect in Section~\ref{sec:lowl}. While the shift is somewhat dependent on mask and frequency, it is present for all masks we consider and not related to the ecliptic masking specifically. The lower panel of Figure~\ref{fig:ecliptic_AL_1d} does, however, illustrate that the residual preference for $A_L>1$ from the $\ell>30$ TT likelihood is weak when the ecliptic equatorial regions are removed. For example, the multifrequency $A_L$ results are consistent with unity within $0.7\sigma$ and $1.0\sigma$ for the \texttt{plik}+\texttt{equa30} and \texttt{plik}+\texttt{equa15} masks.

Given the uncertainties, the shifts in $A_L$ with ecliptic mask are not statistically significant. For the most extreme cuts (left-most and right-most groupings of points in Figure~\ref{fig:ecliptic_AL_1d}, adding \texttt{equa30} and \texttt{pole60} masks) there are no pixels in common. Approximating the posteriors as independent Gaussian distributions\footnote[17]{This is not entirely accurate because the low-multipole likelihoods are used in both results.}, the $A_L$ values differ by $1.4\sigma$ for the 217~GHz and multifrequency fits. Nevertheless, the connection with the Planck scanning strategy and the fact that the data near the ecliptic prefer $A_L>1$ at almost $3\sigma$ motivate exploring the dependence on ecliptic latitude more closely. 

\begin{figure*}
\centering
\includegraphics[width=6.5in]{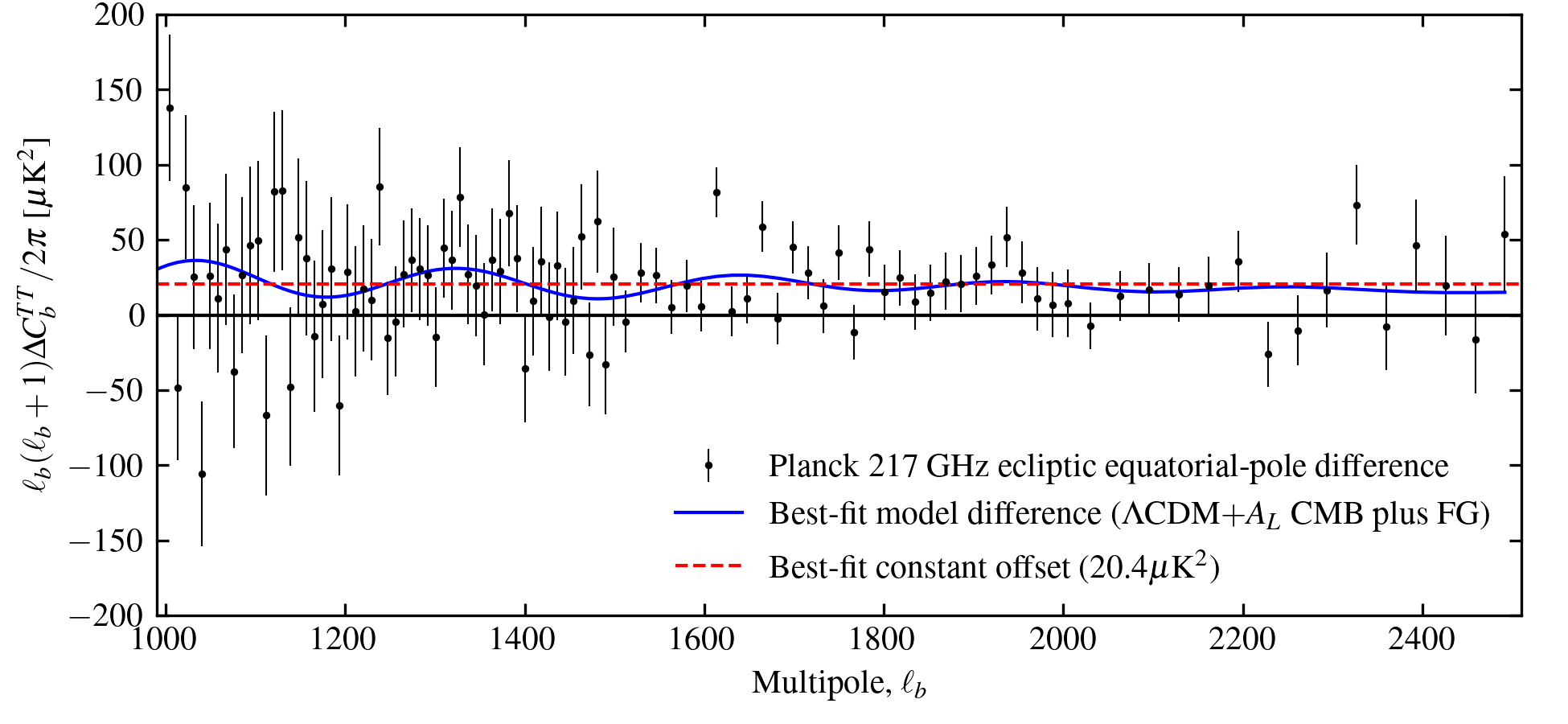}
\caption{Binned angular power spectrum difference at 217~GHz between ecliptic equatorial and polar regions, showing significant excess power near the ecliptic. The difference can be roughly approximated as an offset in $\ell^2C_{\ell}$ (dashed red line), although this shape does not couple significantly to cosmological parameters. The difference in best-fit \lcdm+$A_L$ CMB-plus-foreground models contains additional oscillatory structure, tracing the acoustic peaks (blue line). The marginalized $A_L$ constraints from the two fits are $1.629\pm0.220$ and $1.218\pm0.192$ for the region near the ecliptic and near the ecliptic poles, respectively (Table~\ref{table:ecl_mask}). The masks used correspond to the right-most and left-most Mollweide images in Figure~\ref{fig:ecliptic_AL_1d}.}
\label{fig:ecl_power_diff_217}
\end{figure*}

\subsection{217~GHz Power Spectrum Comparison on Ecliptic Masks}
\label{sec:ecl_power}

Figure~\ref{fig:ecl_power_diff_217} shows the difference between the 217~GHz power spectrum computed near the ecliptic (\texttt{plik}+\texttt{pole60} mask) and near the ecliptic poles (\texttt{plik}+\texttt{equa30} mask). The error bars shown assume the two regions are independent, since there are no pixels in common. There is a significant excess of power near the ecliptic. Modeling this difference as a constant in $D_{\ell}\equiv\ell(\ell+1)C_{\ell}/2\pi$, we find $\Delta D_{\ell}=20.4\pm2.1$~$\mu$K$^2$. The best-fit constant offset is shown as a red dashed line in Figure~\ref{fig:ecl_power_diff_217} and is favored over zero difference at $9.8\sigma$ ($\Delta\chi^2=309.9-213.3=96.6$, with 215 bins and one fitted parameter). We found negligible further improvement from allowing a free power law slope in $\ell$ ($\Delta\chi^2<1$). The difference in the best-fit \lcdm+$A_L$ CMB-plus-FG models from fits to the two regions is shown in blue. Shifts in cosmological parameters, including the shift in $A_L$ discussed above, lead to the oscillatory structure. This matches the difference spectrum better than the constant offset, with a $\chi^2$ of 205.9 compared to 213.3. Similar oscillatory behavior appears in plots in the Planck Collaboration papers comparing fits with different $A_L$ values (e.g., Figure~82 of PL18).

We find that the offset in $\ell^2C_{\ell}$ is largely absorbed into shifts in foreground parameters and does not drive the difference in $A_L$ values on the two masks. Removing this offset by hand from the spectrum computed from the sky near the ecliptic (i.e., subtracting the red line shown in Figure~\ref{fig:ecl_power_diff_217}) yields $A_L=1.579\pm0.210$ in a joint fit with the \texttt{lowl}/\texttt{lowE} likelihoods. This is a difference of only $0.2$ times the original uncertainty (compare with Table~\ref{table:ecl_mask}), ten times smaller than the difference between the low and high ecliptic latitude masks. A scaling in power proportional to $\ell^{-2}$ is flatter than the shape of the Planck Galactic dust templates, which fall off as roughly $\ell^{-2.6}$ (Section~3.3.2 of PL18). Assuming the difference in power shown in Figure~\ref{fig:ecl_power_diff_217} is due to differences in Galactic dust, this suggests a contribution from more compact, point-like structures on small scales.

The $143\times217$~GHz spectrum difference shows qualitatively similar behavior to 217~GHz though less extreme, with a constant offset $\Delta D_{\ell}=7.8\pm1.9$~$\mu$K$^2$ again a leading-order description of the difference, and preferred over zero difference at $4.0\sigma$, but not coupling to the cosmological parameter shifts. There is no significant difference power spectrum difference at 143~GHz, with offset $\Delta D_{\ell}=1.7\pm2.3$~$\mu$K$^2$.

\subsection{Visualizing Residual Galactic Emission Outside the Planck 217~GHz Mask}

While the power spectrum is the direct input to the parameter fitting, it represents a compression of the original sky information, particularly for the anisotropic Galactic emission. Motivated by the results above, we performed a simple map level examination.

\begin{figure*}
\centering
\includegraphics[]{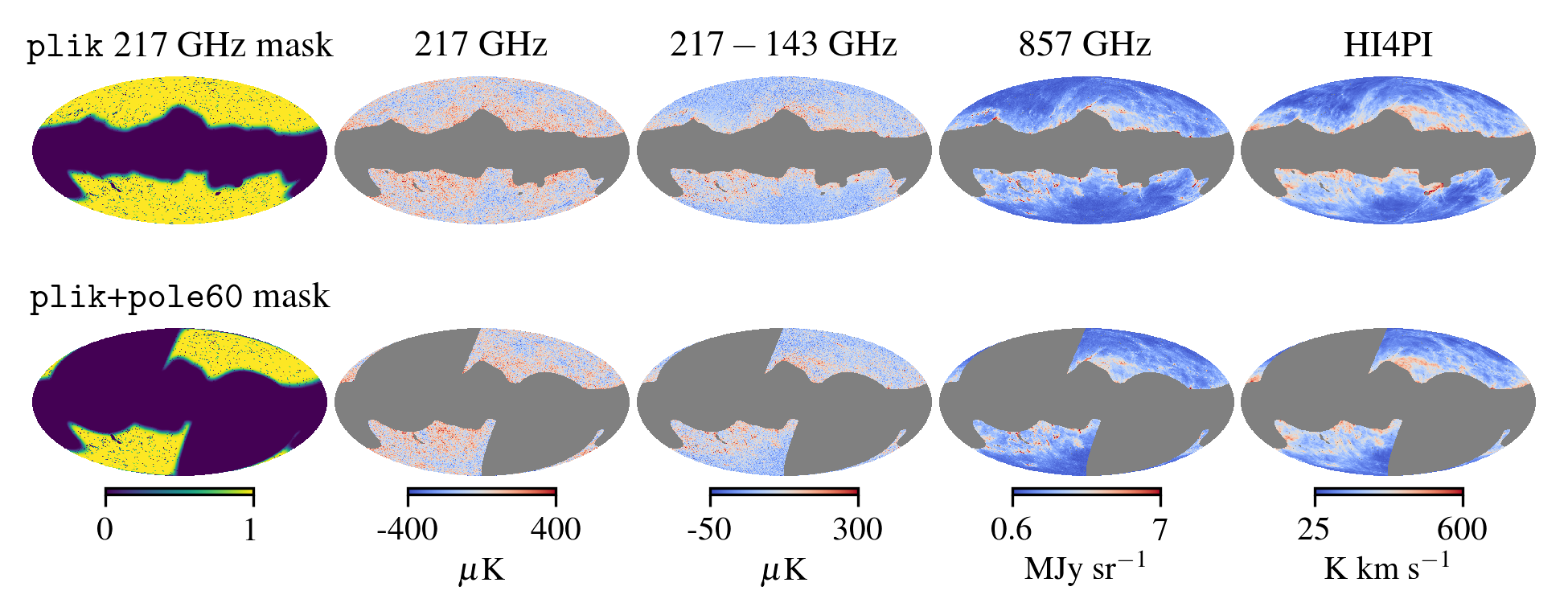}
\caption{Visualization of Galactic structures outside the \texttt{plik} 217~GHz temperature mask. \textsl{(Top row:)} The first image shows the full apodized mask. The other images show a selection of maps with the \texttt{plik} 217~GHz mask applied. Pixels with zero weight in the mask are colored grey. The 217~GHz temperature fluctuations outside the mask are dominated by the CMB, with unmasked Galactic structures more visible in the $217-143$~GHz difference, where the CMB cancels. The 857~GHz Planck data, where the CMB is negligible compared to the dust, and the HI4PI velocity-integrated HI intensity map, show  similar structures to the $217-143$~GHz difference. \textsl{(Bottom row:)} Same as top row, but with additional masking of pixels within $60\degree$ of the ecliptic poles. The Galactic signal is visibly anisotropic, with more bright structures visible near the ecliptic (in the bottom left and top right quadrants of the images) than near the ecliptic poles.}
\label{fig:ecliptic_masks}
\end{figure*}

Figure~\ref{fig:ecliptic_masks} shows the \texttt{plik} 217~GHz mask, as well as the product of this mask with our \texttt{pole60} mask removing pixels within $60\degree$ of the ecliptic poles. We show four data maps with these two masks applied: (1) the 217~GHz full-mission temperature map, (2) the pixel-wise difference between the 217~GHz and 143~GHz full-mission maps, (3) the 857~GHz full-mission intensity map, and (4) the HI4PI velocity-integrated HI intensity map \citep{HI4PI:2016}. Large-scale CMB fluctuations hamper visual identification of bright dust regions admitted by the \texttt{plik} mask at 217~GHz, although an excess in the lower left quadrant is still apparent by eye. The other three maps, where the CMB is negligible, show similar bright degree-scale Galactic structures. The HI4PI map is included as an external dataset comparison. There is a well-established correlation between the HI and sub-millimeter/microwave Galactic thermal dust emission for low column densities \citep[e.g.,][]{planck2011-6.6,lenz/dore/lagache:2019}. Visually, the features in the HI4PI map are similar to those in the 857~GHz map, although there are some differences, including the bright streak between the Magellanic clouds.

Overall, the brightest Galactic features that are admitted by the \texttt{plik} mask are predominantly at low ecliptic latitudes. This is consistent with the excess of power measured at 217~GHz near the ecliptic (Section~\ref{sec:ecl_power}). While our simple test removing a constant offset in $\ell^2C_{\ell}$ did not greatly impact $A_L$, this nevertheless raises the possibility that the shifts in $A_L$ with ecliptic mask may be somehow related to unmasked Galactic emission, as opposed to arising from some physical or observational effect more directly connected to ecliptic latitude.

\section{Additional Sky Cuts Based on 857~GHz Dust Intensity}
\label{sec:dust}

In this Section we investigate additional masking targeting bright Galactic dust structures, how this impacts $A_L$, and whether this explains the ecliptic latitude dependence in the previous Section.

The Planck Galactic masks were constructed using a CMB-subtracted 353~GHz intensity map. For a given intensity threshold, the initial binary mask was then coarsely smoothed with a $10\degree$ Gaussian before $\sigma=2\degree$ apodization was applied to pixels at the boundary \citep[Appendix~A of][]{planck2014-a11}. Here we instead use thresholding of the 857~GHz full-mission intensity map, where the CMB is negligible compared to the dust. We forego any coarse smoothing step and apply apodization directly to the binary mask corresponding to the specified intensity threshold, using $\sigma=30'$ smoothing. This approach has the potential disadvantage of increasing the small-scale power in the mask, although based on simulations this does not invalidate our covariance matrix approximations (Appendix~\ref{app:sims}), and it has the advantage of removing the bright dust structures at finer scales than in the \texttt{plik} mask construction.

\subsection{Results from Additional Dust Masking at 217~GHz}

\begin{figure*}
\centering
\includegraphics[width=6.5in]{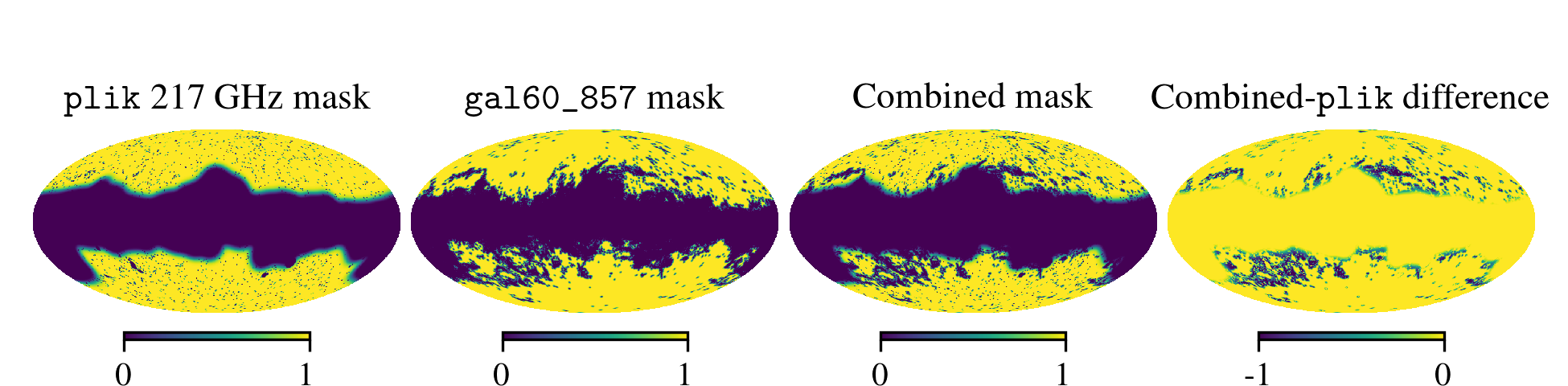}
\caption{Comparison of original Planck 2018 \texttt{plik} likelihood mask for 217~GHz with our \texttt{gal60\_857} mask constructed using thresholding in 857~GHz intensity. We perform analysis with the combination (product) of the two masks, which retains sky fraction $f_{\rm sky}=0.42$, compared to 0.50 for the original \texttt{plik} mask. The additional masked regions, highlighted in the combined-\texttt{plik} mask difference, remove Galactic dust structures on finer angular scales than the $10\degree$ smoothing scale used in the \texttt{plik} mask construction, particularly in the lower left and upper right portions of the image.}
\label{fig:mask857}
\end{figure*}

We used an 857~GHz intensity cut retaining a sky fraction $f_{\rm sky}=0.60$, chosen to match the sky fraction of the original \texttt{plik} 217~GHz Galactic mask, prior to apodization. We label this mask `\texttt{gal60\_857}'. We discuss results with different 857~GHz intensity cuts in Section~\ref{sec:multifreq}, below. We constructed new 217~GHz HM masks by multiplying the \texttt{plik} 217~GHz HM masks by the \texttt{gal60\_857} mask, shown visually in Figure~\ref{fig:mask857}. The final combined masks have $f_{\rm sky}=0.42$, compared to 0.50 for the \texttt{plik} 217~GHz HM masks. The difference here is modest because most of the pixels masked in the \texttt{gal60\_857} mask were already masked in the \texttt{plik} masks.

We constructed new likelihoods using the new 217~GHz masks, following the procedure described in Section~\ref{sec:computations}. The additional masking impacts the $143\times217$~GHz spectrum, as well as the 217~GHz spectrum, however the 100 and 143~GHz masks and spectra are unchanged. 

Table~\ref{table:gal60_857} shows 1D marginalized constraints on $A_L$ with and without the additional masking. The $A_L$ value at 217~GHz changes from $1.362\pm0.147$ to $1.182\pm0.151$, a decrease by more than the original uncertainty, with only a small change in the error bar. A decrease by more than half the original uncertainty is also seen at $143\times217$~GHz. Notably, however, the multifrequency $A_L$ result, with the 100 and 143~GHz spectra included, does not shift by any meaningful amount. We first examine the significance of the 217~GHz shift, before turning to what happens in the multifrequency case.

\begin{table}
  \centering
  \caption{Impact of Additional Dust Masking at 217~GHz on $A_L$ Constraints}
  \begin{tabular}{lcc}
\hline
TT Spectrum&$A_L$&$A_L$\\
&(\texttt{plik} masks)&(+\texttt{gal60\_857})\\
\hline
\hline
100~GHz&$1.175\pm0.185$&unchanged\\
143~GHz&$1.171\pm0.111$&unchanged\\
$143\times217$~GHz&$1.201\pm0.114$&$1.123\pm0.117$\\
217~GHz&$1.362\pm0.147$&$1.182\pm0.151$\\
All&$1.240\pm0.094$&$1.244\pm0.096$\\
\hline
\end{tabular}
\label{table:gal60_857}
\end{table}

\subsection{Assessment of $A_L$ Shift Significance at 217~GHz}
\label{sec:ALshift217}

To quantitatively assess the  significance of the shifts in $A_L$ we need to account for the fact that the constraints are highly correlated (all the pixels used in the new mask were also used in the \texttt{plik} mask). The Planck Collaboration used a ``difference-of-covariance'' method \citep{gratton/challinor:2020} to estimate the significance of parameter shifts between MCMC fits when one uses a subset of the data in the other (Section~3.9.1 of PL18). The covariance of the parameter shift vector is approximated as the difference of the parameter covariance matrices. Under certain conditions, essentially when the posterior uncertainties are well described by the Fisher matrix, this is an accurate and fast method, requiring only the posterior covariance estimated from the chains and removing the need for any expensive simulations. Based on this method, the significance of the 217~GHz $A_L$ shift from adding the \texttt{gal60\_857} masking is $5.2\sigma$ (i.e., incompatible with simply being a statistical fluctuation). For the $143\times217$~GHz spectrum alone, the corresponding significance is $3.0\sigma$. 

By performing \lcdm$+A_L$ MCMC fits to a subset of the simulations described in Appendix~\ref{app:sims} we found that the difference-of-covariance method is not always accurate for assessing $A_L$ shift significance between masks. The main issue is that the parameter precision does not monotonically decrease as more of the sky is masked, and there is a significant correlation between the posterior $A_L$ mean and uncertainty.\footnote[18]{This effect is not unique to $A_L$, and exists for the \lcdm\ parameters as well, however we will focus on $A_L$ here.} The simulations were generated with cosmological parameters matching the Planck 2018 best-fit \lcdm\ model, with $A_L=1$. Realizations that produce high $A_L$ values also tend to have wider $A_L$ posteriors, despite the fact that departures from $A_L=1$ in the simulations are due solely to statistical fluctuations. We have verified that this is not merely a result of numerical issues due, for example, to the finite number of steps in converged MCMC runs. Instead it is likely a consequence of a breakdown of the validity of the Fisher matrix approximation for the posterior covariance. That is, the posterior covariance is not always well approximated as the Fisher matrix evaluated at the true input cosmology, but generally requires some next-order correction that depends on the scatter of the sky modes in each realization.

Based on shifts in $A_L$ from adding the \texttt{gal60\_857} mask in 100 simulations, we estimate the significance of the shift in the real 217~GHz data at $3.1-3.5\sigma$. None of the simulations had $A_L$ as high as the real data, however we found that the magnitude and direction of the $A_L$ shifts in the simulations did not correlate with the preferred $A_L$ values. We therefore fit a Gaussian to the distribution of $A_L$ shifts in the simulations, and used this to quantify the likelihood of the shift seen in the data. The range quoted comes from propagating uncertainty in the Gaussian width due to the limited number of simulations. Despite this uncertainty, there is nevertheless evidence that the shift in $A_L$ is connected to the dust foreground emission removed by the \texttt{gal60\_857} mask, rather than arising solely from statistical fluctuations. In light of the simulations we regard the $5.2\sigma$ difference-of-covariance result above as overestimating the shift significance, due to the reduction in the preferred $A_L$ value being accompanied by a decrease in the posterior width, and this partially canceling the increase in uncertainty expected from the reduced sky fraction alone.

We have highlighted the limitation of the difference-of-covariance approximation here partly because of its now-widespread use by teams beyond Planck, including ACT \citep[e.g.,][]{choi/etal:2020} and SPT \citep[e.g.,][]{dutcher/etal:2021}. Unfortunately there is no obvious indication that the approximation is breaking down, except potentially in extreme cases where a decrease in the amount of data used actually leads to a tightening in the posterior width, or vice versa. The recommendation then is to perform simulations to check the difference-of-covariance validity, particularly for cases where the data seem to deviate from statistical expectations.

\subsection{Multifrequency Results with Additional 217~GHz Dust Masking}
\label{sec:multifreq}

\begin{figure*}
\centering
\includegraphics[width=6.5in]{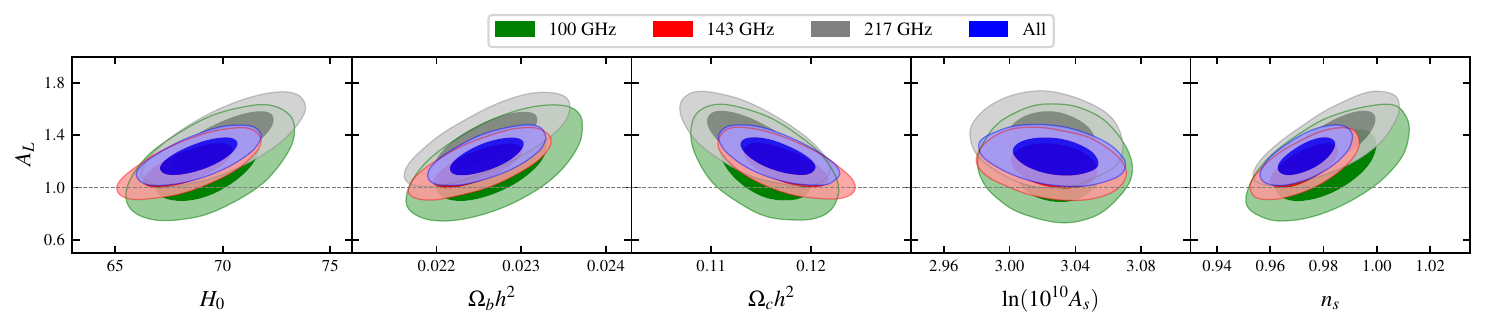}
\includegraphics[width=6.5in]{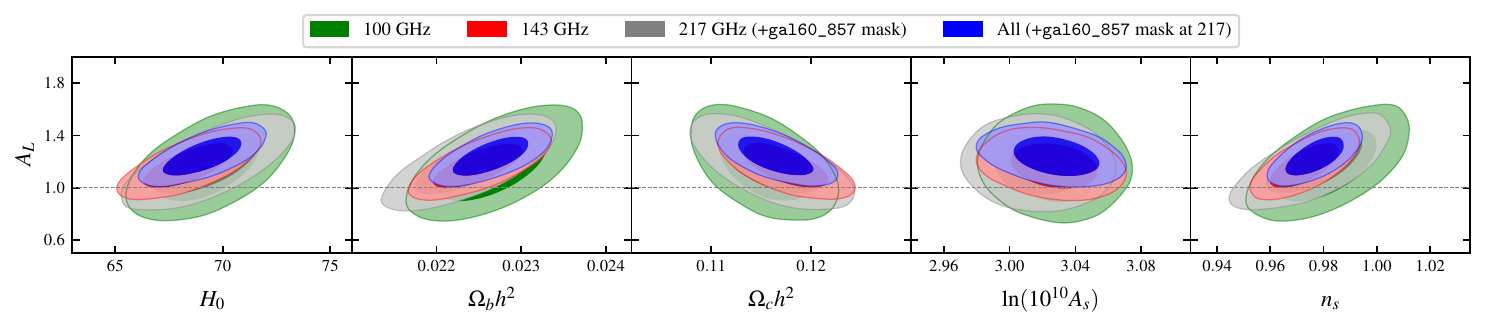}
\includegraphics[width=6.5in]{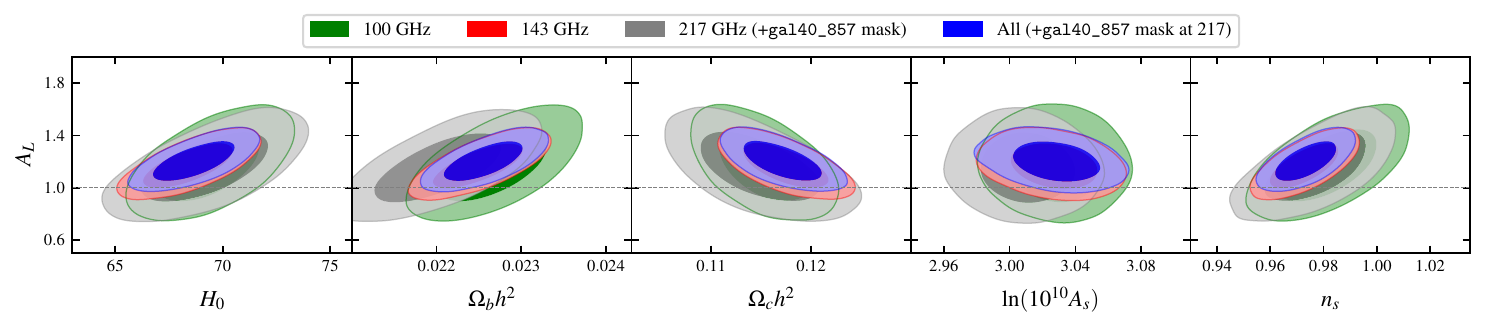}
\caption{Correlations between $A_L$ and \lcdm\ parameters for single frequency and multifrequency fits, with the original Planck masks (top row), with the \texttt{gal60\_857} dust mask added at 217~GHz (middle row), and with the more stringent \texttt{gal40\_857} mask added at 217~GHz (bottom row). We show the 2D correlations to highlight that the degeneracy directions differ slightly with frequency and that the multifrequency $A_L$ constraint is not simply the average of the constraints from the individual frequencies. The preference for higher $A_L$ values at 217~GHz is reduced with the \texttt{gal60\_857} mask. This downward shift at 217~GHz is inconsistent with a statistical fluctuation (Section~\ref{sec:ALshift217}) and indicates some connection between $A_L$ at 217~GHz and Galactic dust masking. The preference for $A_L>1$ remains at $2.5\sigma$ in the multifrequency fit for the \texttt{gal60\_857} mask, but decreases as further 217~GHz masking is applied (to $2.1\sigma$ in the bottom row with the \texttt{gal40\_857} mask, or $1.8\sigma$ with 217~GHz removed entirely). The 100 and 143~GHz contours are the same in each row and the \texttt{lowl} and \texttt{lowE} $\ell<30$ likelihoods are included in each case.}
\label{fig:AL_single_multifreq}
\end{figure*}

Despite the significant impact of the \texttt{gal60\_857} mask on $A_L$ constraints from 217~GHz, Table~\ref{table:gal60_857} shows that the effect on the multifrequency constraint is negligible. With the additional masking, none of the individual TT spectra have an $A_L$ posterior mean greater than $1.2$, yet the combined constraint is centered at $1.244$. 

Figure~\ref{fig:AL_single_multifreq} shows 2D posteriors for $A_L$ and the \lcdm\ parameters for the single frequency and multifrequency fits for several masks. While all the individual frequencies are consistent with $A_L=1$ within around $1.5\sigma$, the slightly different degeneracy directions mean the combined $A_L$ constraint is not simply equal to a weighted average of the single frequency constraints. For the \texttt{gal60\_857} case (middle row), one can see for example in the $\Omega_bh^2$ panel that the overlap between the different contours is more complete for higher $\Omega_bh^2$ values, where $A_L$ is also higher.

\begin{table*}
  \centering
  \caption{$A_L$ Constraints from Combining Additional Dust Masking at 217~GHz with Ecliptic Mask Removing Pixels within $60\degree$ of the Ecliptic Poles}
  \begin{tabular}{lccc}
\hline
TT Spectrum&$A_L$&$A_L$&$A_L$\\
&(\texttt{plik}+\texttt{pole60})&(+\texttt{gal60\_857})&(+\texttt{gal40\_857})\\
\hline
\hline
100~GHz&$1.33\pm0.26$&unchanged&unchanged\\
143~GHz&$1.23\pm0.16$&unchanged&unchanged\\
$143\times217$~GHz&$1.38\pm0.16$&$1.39\pm0.17$&$1.48\pm0.23$\\
217~GHz&$1.63\pm0.22$&$1.52\pm0.23$&$1.67\pm0.32$\\
All&$1.41\pm0.14$&$1.40\pm0.15$&$1.37\pm0.15$\\
\hline
\end{tabular}
\label{table:pole60_gal60_857}
\end{table*}

While adding the \texttt{gal60\_857} mask at 217~GHz did not impact the combined $A_L$ constraint, it is important to note that the 217~GHz data do still contribute to the overall preference for $A_L>1$. Removing the 217~GHz data entirely (so using only the 100 and 143~GHz spectra) yields $A_L=1.198\pm0.109$ in conjunction with the \texttt{lowl}/\texttt{lowE} likelihoods. Further masking at 217~GHz (further downweighting the contribution from the 217~GHz data) should shift $A_L$ towards this result and we verified that this is indeed the case. For example, choosing a stricter intensity threshold at 857~GHz to retain only 40\% of the sky, combining this \texttt{gal40\_857} mask with the \texttt{plik} 217~GHz masks, and repeating all calculations, we find $A_L=1.206\pm0.100$. The $A_L$ constraint from 217~GHz only is fairly robust to this additional masking, with the posterior mean remaining in the range $1.15-1.20$ even though the uncertainty relative to the \texttt{plik}+\texttt{gal60\_857} case increases by as much as a factor of 1.5 due to the reduced sky area. As shown in the bottom row of Figure~\ref{fig:AL_single_multifreq}, there are some shifts in the \lcdm\ parameters and their uncertainties are also significantly increased.

We investigated raising the 857~GHz intensity threshold, still combining with the \texttt{plik} mask, but removing less additional sky, to test whether the 217~GHz $A_L$ shifts could be connected with a specific change in the threshold. For masks \texttt{gal70\_857}, and \texttt{gal65\_857}, the posterior $A_L$ means are 1.32 and 1.25, indicating that the downwards shifts in $A_L$ at 217~GHz arise between the \texttt{gal70} to \texttt{gal60} masking. We have not performed a more in-depth analysis of exactly which features on the sky drive the mask-related shifts in $A_L$ at 217~GHz.

We also investigated applying additional dust masking to all the frequency maps (including 100 and 143~GHz as well as 217), using the thresholded 857~GHz intensity map as described above. We found no clear trend between $A_L$ and sky area for fits with only the 100 or 143~GHz data or for the associated multifrequency fits. Relative to the 143~GHz \texttt{plik} mask, masking up to \texttt{gal40\_857} increases parameter uncertainties by around a factor of two, however the $A_L$ posteriors continue to lie in the range $1.15-1.20$. The 100~GHz constraints, which do not use multipoles $\ell>1200$, are always weaker, but similarly show no particular trend with additional masking.

Summarizing the results above, we have presented evidence that the constraint on $A_L$ from 217~GHz alone is biased high on the \texttt{plik} mask due to some issue connected with the residual Galactic dust. When enough additional masking is applied, the posterior $A_L$ values from 217~GHz shift from around 1.36 to around 1.2, more closely matching values obtained from the other frequencies. In the multifrequency fit, the high values of $A_L$ preferred from 217~GHz alone are disfavored or ruled out by the other frequencies, where the residual dust contribution is significantly smaller. Applying the \texttt{gal60\_857} mask at 217~GHz does not impact $A_L$ in the multifrequency case, and the preference for $A_L>1$ remains at $2.5\sigma$ ($A_L=1.244\pm0.096$). Applying additional dust masking at 217~GHz does reduce the preference for high $A_L$ in the multifrequency fit, and removing 217~GHz altogether leads to a $1.8\sigma$ preference for $A_L>1$ ($1.198\pm0.109$).

\subsection{Does Additional Galactic Dust Masking Impact the Dependence of $A_L$ on Ecliptic Latitude?}
\label{sec:eclplusdust}

To directly test the extent to which dust masking can explain the ecliptic latitude dependence discussed in Section~\ref{sec:ecliptic} we applied additional dust masking to the mask retaining only pixels within $30\degree$ of the ecliptic plane (i.e., masking $60\degree$ around the poles). Table~\ref{table:pole60_gal60_857} shows $A_L$ results from the original ecliptic mask and with adding the \texttt{gal60\_857} and \texttt{gal40\_857} masks. Despite the increased uncertainties, particularly for the \texttt{gal40\_857} mask, there is no reduced preference for $A_L>1$, or coherent downward shift in the mean values, even for the $143\times217$ or 217~GHz spectra individually.

Based on these results, the preference for $A_L>1$ at low ecliptic latitudes is not predominantly tied to regions with bright 857~GHz dust emission. There appear to be at least two separate mechanisms contributing to the shifts in $A_L$ with masking at 217~GHz.

\section{Impact of the Low-Multipole Temperature Power Spectrum on $A_L$}
\label{sec:lowl}

This work focuses on the high-multipole Planck data and the impact of analysis choices like the masking and frequency selection. The $\ell<30$ \texttt{lowl} TT data also play a  role in the $A_L$ anomaly, indirectly contributing to the preference for $A_L>1$, and we now examine this in more detail.

\begin{figure*}
\centering
\includegraphics[width=6.5in]{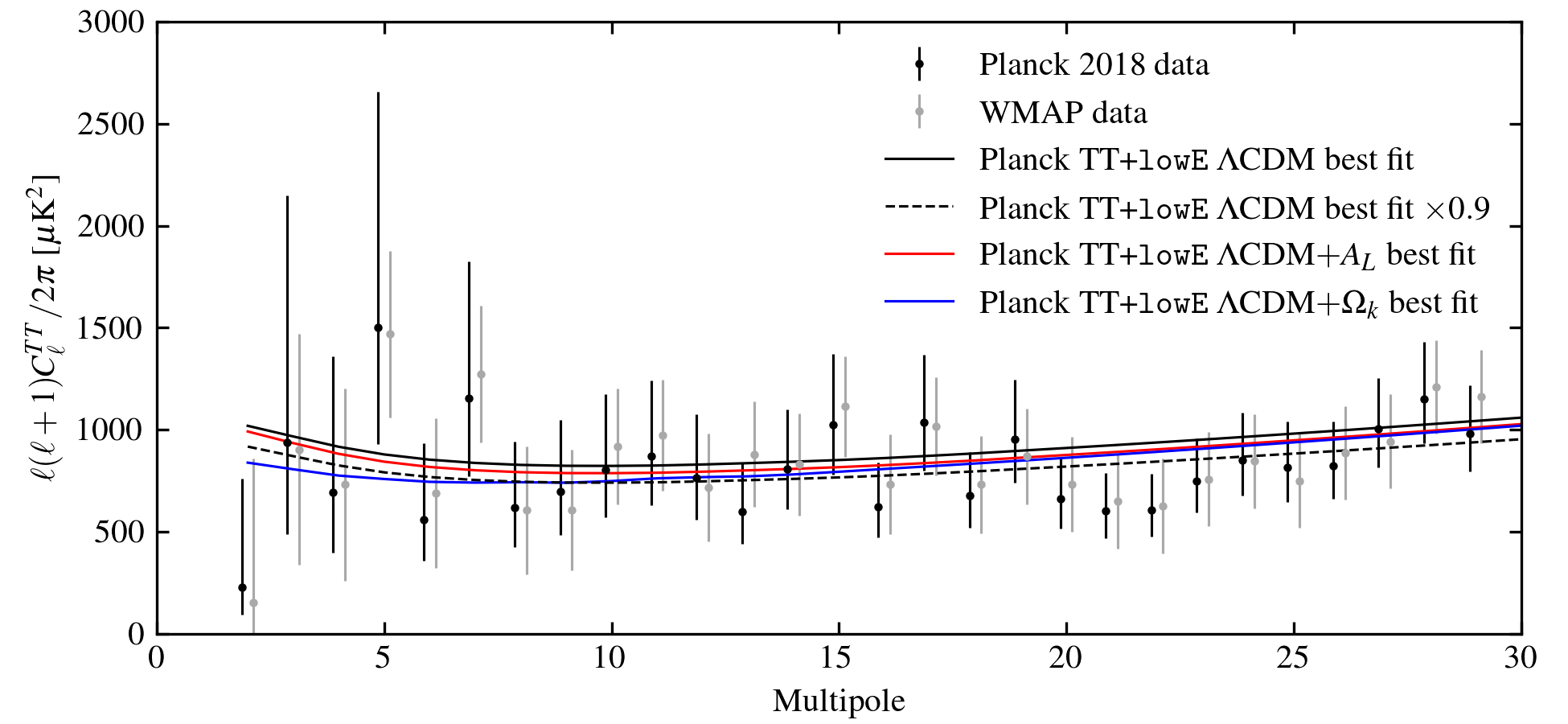}
\caption{Comparison of low-multipole temperature power spectrum measurements and model fits. The \lcdm\ fit to all the Planck TT data is high relative to the $\ell<30$ data, particularly at $20\leq\ell\leq22$. Applying a multiplicative rescaling factor of 0.9 to this model provides the best match to $\ell<30$ ($\Delta\chi^2=-4.0$ w.r.t \lcdm). Lensing does not directly impact $\ell<30$, but varying $A_L$ allows the \lcdm\ parameters to shift and partially improve the fit here ($\Delta\chi^2=-2.3$). The Planck $\ell<30$ TT data contribute $0.4\sigma$ of the preference for $A_L>1$. Varying $\Omega_k$ instead of $A_L$ slightly improves the fit at $\ell<20$, with $\Delta\chi^2=-2.6$. This is due to a suppression of the late-time ISW effect. Using WMAP TT instead of Planck at $\ell<30$ reduces the preference for $A_L>1$ by $0.2\sigma$.}
\label{fig:lowl}
\end{figure*}

Figure~\ref{fig:lowl} shows Planck 2018 and WMAP TT measurements at $\ell<30$ along with several model fits. In the \lcdm\ fit to all the Planck TT data, the model is high of most of the $\ell<30$ data, particularly at $20\leq\ell\leq22$ (solid black line). By eye one can see that the $\ell<30$ data would prefer a lower overall amplitude, however there is insufficient freedom to achieve this in \lcdm\ without some more severe worsening of the fit at higher multipoles. Applying a scale-independent multiplicative rescaling to the \lcdm\ fit, we find that a factor 0.9 best matches the $\ell<30$ Planck data (dashed black line), with an improved $\chi^2$ relative to the unscaled fit of $\Delta\chi^2=19.6-23.6=-4.0$.\footnote[19]{The likelihood for the $\ell<30$ power spectrum multipoles is non-Gausssian due to the small number of modes, however a transformation was applied to correct for this in the Planck \texttt{lowl} likelihood, such that $-2$ times the log-likelihood can be interpreted as a standard $\chi^2$ (see, e.g., Section~2.1 of PL18).} This still does not perfectly match the points at $\ell=21,22$, although lowering the power further pays too large a penalty from other multipoles.

As mentioned in Section~\ref{sec:ecliptic}, varying $A_L$ does not directly impact $\ell<30$, because the lensing effect is negligible on these scales. In the \lcdm+$A_L$ fit, however, there is sufficient freedom to achieve a somewhat better match to $\ell<30$ while also improving the fit at higher multipoles, via partial degeneracies between $A_L$ and parameters including $A_s$ and $n_s$ (red line in Figure~\ref{fig:lowl}). For \lcdm+$A_L$, the fit to the $\ell<30$ TT data is improved by $\Delta\chi^2=-2.3$ over \lcdm\ (this shift is quoted for the best-fit $A_L$ value of 1.26). Varying $A_L$ in the fit to the full multipole range therefore provides a little over half of the $\chi^2$ improvement in principle possible at $\ell<30$ (comparing to the result from multiplicatively rescaling the \lcdm\ theory curve).

Varying $\Omega_k$ instead of $A_L$ produces similar results for the TT spectrum at $\ell\gtrsim20$ but allows for less power at $\ell<20$ and a slightly improved overall fit to the Planck low-multipole likelihood, with $\Delta\chi^2=-2.6$ over \lcdm\ (blue line in Figure~\ref{fig:lowl}). This improvement was also noted by \cite{planck2016-l06} and arises because allowing a negative $\Omega_k$ (along with shifts in \lcdm\ parameters required to maintain a good match to the acoustic peak structure) leads to a reduction in the late-time integrated Sachs-Wolfe (ISW) effect. Specifically, $\Omega_m$ is higher, and $\Omega_{\Lambda}$ is lower, relative to \lcdm. This does not happen when varying $A_L$ and the improvement at low multipoles in that case comes from a reduction of primordial power through shifts in $A_s$ and $n_s$.

A deficit of power at low multipoles was first measured by WMAP and has been widely discussed in the context of  anomalous features in the large-scale CMB temperature fluctuations \citep[e.g.,][]{bennett/etal:2011,planck2013-p09,schwarz/etal:2016}. In fact, the agreement between the WMAP and Planck points in Figure~\ref{fig:lowl} is not perfect, even though instrument noise is negligible on these scales.\footnote[20]{The values and uncertainties in Figure~\ref{fig:lowl} were taken directly from the power spectrum files released by the WMAP and Planck Collaborations, found at \url{https://lambda.gsfc.nasa.gov/product/wmap/dr5/pow\_tt\_spec\_get.html} and \url{http://pla.esac.esa.int/pla/\#cosmology}, respectively. We have not attempted to homogenize treatment of cosmic variance or other effects.} The masking and foreground treatment are different, and this may explain some or all of the difference (see discussion in Section~2.1 of PL18, including direct comparison between Planck results on different masks).

We performed fits replacing the Planck \texttt{lowl} likelihood with the corresponding $2\leq\ell<30$ WMAP likelihood (but not using any other WMAP temperature or polarization data). For the multifrequency fit on the original \texttt{plik} masks, this reduced the preference for $A_L>1$ by around $0.2\sigma$, from $2.5\sigma$ to $2.3\sigma$, with the $A_L$ posterior shifting from $1.240\pm0.094$ to $1.223\pm0.095$. When $\ell<30$ TT data are removed altogether the preference for $A_L>1$ is $2.1\sigma$ ($1.205\pm0.096$). In detail, then, the WMAP and Planck $\ell<30$ TT data do have a different impact on the $A_L$ results, with the WMAP data only favoring higher $A_L$ values over \lcdm\ at roughly half the significance of Planck. While we quote only the $A_L$ posteriors here the \lcdm\ parameters are partially correlated with $A_L$ and similarly shift at around the $0.2\sigma$ level when the WMAP $\ell<30$ likelihood is used instead of Planck. We also investigated using the Planck 2015 $\ell<30$ likelihood in place of that from 2018, but found that this had a smaller impact ($<0.1\sigma$ in $A_L$ and other parameters).

Overall the choice of $\ell<30$ TT likelihood is a subdominant contributor to the $A_L$ anomaly, which is primarily driven by the higher multipoles. We do, however, view this difference between WMAP and Planck as motivation for revisiting the low-$\ell$ analysis choices (e.g., masking and foreground modeling), since these experiments should provide near-identical results for these temperature modes. We note here that the higher-multipole WMAP and Planck TT spectrum measurements, where power-spectrum-based rather than pixel-based likelihoods are used, agree within statistical expectations when masking differences and common cosmic variance are accounted for \citep{huang/etal:2018}.

\section{Additional Tests for $A_L$ Shifts}
\label{sec:additional}

In this Section we report two further tests designed to potentially shed light on the preference for $A_L>1$ in the Planck temperature data, particularly at low ecliptic latitudes. Unlike in Sections~\ref{sec:ecliptic} and \ref{sec:dust}, these tests did not result in notable shifts in $A_L$. We describe them here regardless since they limit or rule out plausible systematic origins of the $A_L$ anomaly and may inform future investigations.

\begin{figure}
\centering
\includegraphics[]{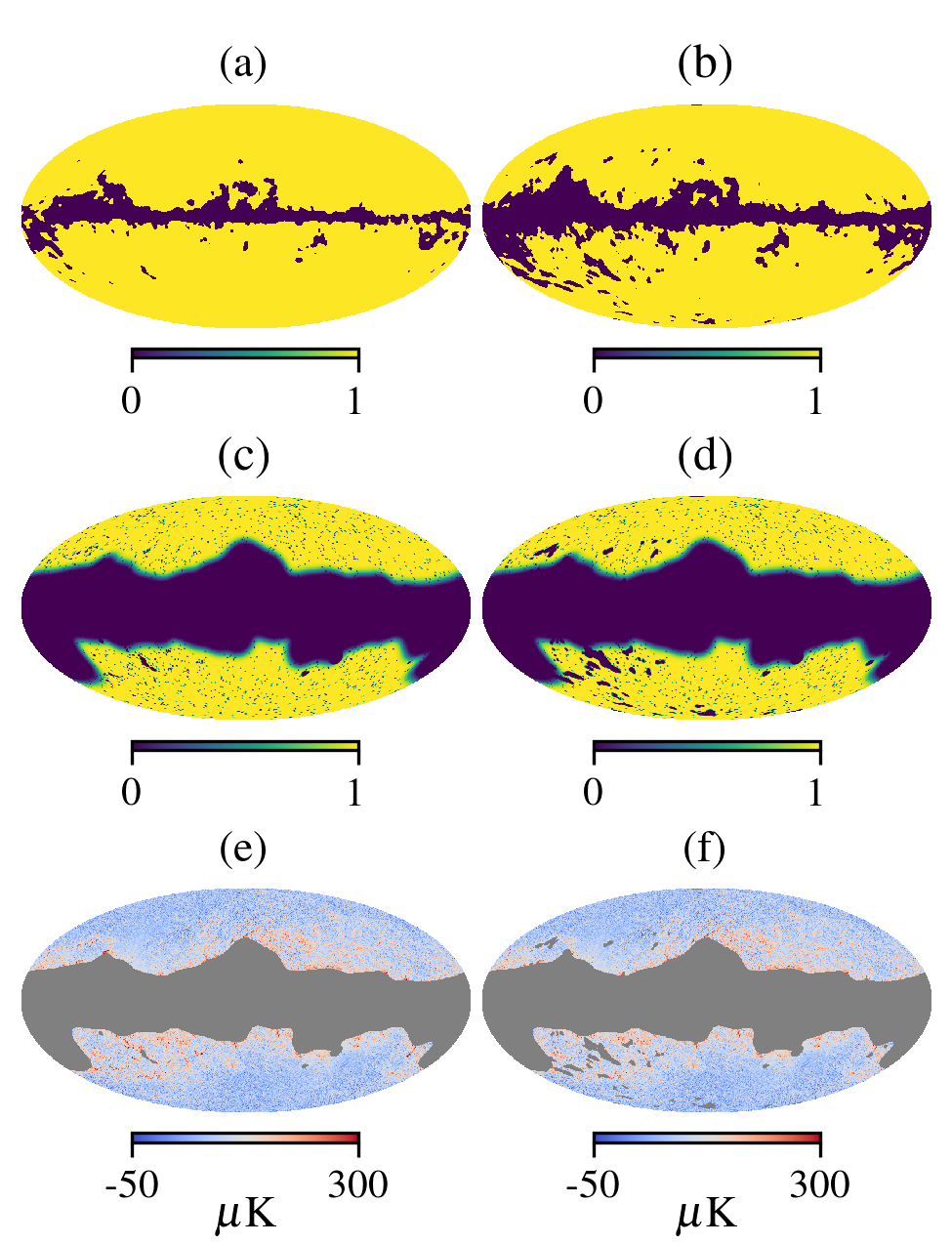}
\caption{Comparing 217~GHz sky masks for default and aggressive CO line intensity masking thresholds. (a) Binary CO mask removing pixels above 1~K$_{\rm RJ}$~km~s$^{-1}$, as used in the \texttt{plik} mask construction. (b) Our more aggressive CO mask thresholded at 0.1~K$_{\rm RJ}$~km~s$^{-1}$. (c) Original \texttt{plik} 217~GHz mask. (d) \texttt{plik} 217~GHz mask multiplied by more aggressive CO mask. (e) $217-143$~GHz map difference with masked pixels in \texttt{plik} mask greyed out. (f) $217-143$~GHz map difference with more aggressive CO mask.}
\label{fig:CO_mask}
\end{figure}

\subsection{Additional CO Masking}

We investigated whether residual CO line emission in the 217~GHz data could be impacting the $A_L$ results. The \texttt{plik} masks at 100 and 217~GHz include removal of pixels with CO line intensity greater than 1~K$_{\rm RJ}$~km~s$^{-1}$ based on the `Type-3' CO map produced in the 2013 Planck release \citep[see Appendix~A of][for details]{planck2014-a13}. No residual CO component was then included in the \texttt{plik} foreground model used for the cosmological fitting. The lack of full-sky, high-resolution CO line intensity maps unfortunately prevents masking CO regions more aggressively based on external data. We therefore followed the Planck Collaboration approach, thresholding a smoothed version of the Planck 2013 Type-3 CO map, but adopted a threshold ten times more stringent, removing pixels with CO line intensity greater than 0.1~K$_{\rm RJ}$~km~s$^{-1}$. Noise in the CO map prevents pushing below this limit.

The original and more aggressive CO masks are shown in Figure~\ref{fig:CO_mask}. While many of the additional masked pixels were already assigned zero weight in the \texttt{plik} 217~GHz mask (largely because of the Galactic dust masking), a number of degree-scale regions, predominantly in the lower left quadrant (in Galactic coordinates) are removed by the stricter CO cut. There is partial overlap between these regions and those masked by the 857~GHz intensity thresholding in Section~\ref{sec:dust} (see Figure~\ref{fig:ecliptic_masks}).

We found that the stricter CO mask has only a small effect on $A_L$ from 217~GHz, shifting the marginalized constraint from $A_L=1.362\pm0.147$ to $1.335\pm0.144$, a difference of around 0.2 times the uncertainty. There is a measurable decrease in 217~GHz power from the additional masking, roughly constant in $\ell^2C_{\ell}$, similar to that found in Section~\ref{sec:ecl_power} and shown in Figure~\ref{fig:ecl_power_diff_217}, however this is largely absorbed into the foreground parameters. We do not interpret this shift in power as necessarily indicating a significant CO residual, however, since the regions with bright CO also tend to have bright dust emission.

Given the small shift in $A_L$ at 217~GHz we did not perform a fit with other frequencies (note that the 143~GHz band does not overlap the CO lines) and conclude that residual CO is not a significant contributor to the $A_L$ anomaly.

\subsection{Variation of Planck 1D Effective Beam Window Functions with Ecliptic Mask}

As discussed earlier, the Planck scan coverage was highly inhomogeneous. While the variation in hits per pixel is accounted for in our covariance matrix calculations (and updated with changes in the masking), we have assumed to this point that the effective 1D beam window function, $b_{\ell}$, used to correct the measured power spectrum and obtain an unbiased sky spectrum estimate, is independent of sky region. The Planck beams are asymmetric \citep{planck2013-p03c} and this effect would not be averaged out (``symmetrized'') for pixels near the ecliptic plane that were always scanned at nearly the same orientation. We therefore performed a test to assess whether a failure to account for some dependence of the effective beam function on ecliptic masking could explain or contribute to the shifts we found in $A_L$.

\begin{figure}
\centering
\includegraphics[width=3.25in]{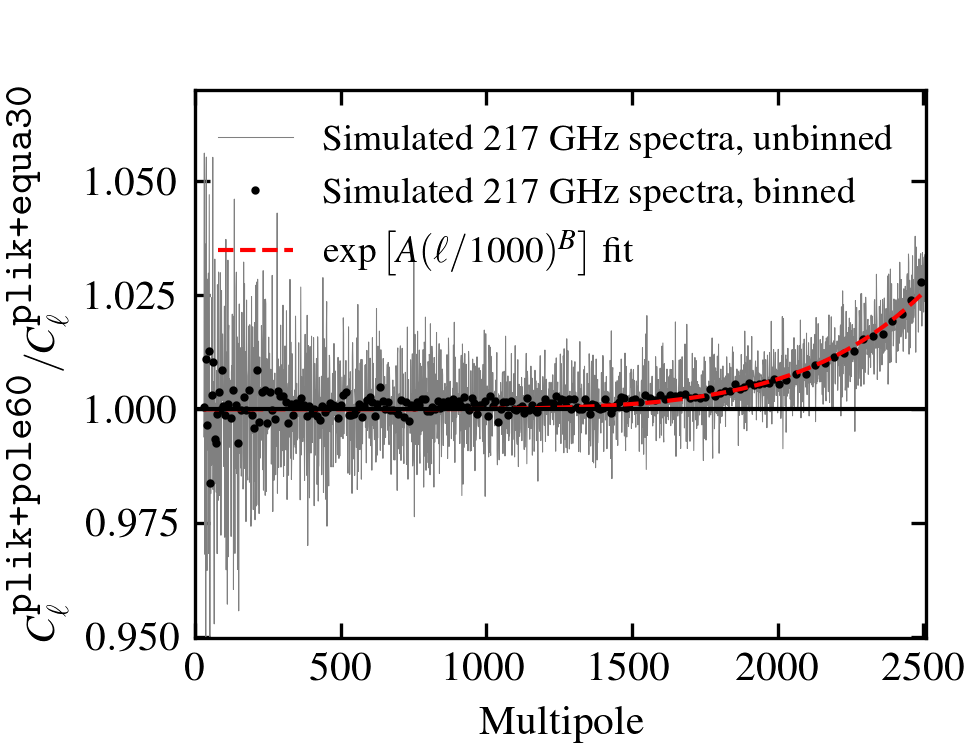}
\caption{Ratio of power from low ecliptic latitudes (masking around the ecliptic poles) to power from high ecliptic latitudes (masking around the ecliptic plane), estimated using 1000 Planck Full Focal Plane (FFP10) simulations at 217~GHz. We interpret the deviation from unity at $\ell>1500$ as a difference in effective Planck beam function on the different sky regions due to beam asymmetries and the scan pattern. As discussed in the text, this shape does not couple significantly to $A_L$ or other cosmological parameters, and does not explain the shifts in $A_L$ with ecliptic latitude.}
\label{fig:sim_power_ratio}
\end{figure}

We computed angular power spectra from the 1000 FFP10 CMB simulations made available by the Planck Collaboration with the 2018 release on masks using sky at low and high ecliptic latitudes (with a split $30\degree$ from the plane as described in Section~\ref{sec:ecliptic}). Figure~\ref{fig:sim_power_ratio} shows the ratio of the mean power averaged over the realizations. We used the 217~GHz CMB-only simulations (without noise or other sky signals), since the 217~GHz data exhibit the strongest dependence on ecliptic masking, and performed mask deconvolution using the mode-coupling matrices appropriate for the \texttt{plik}+\texttt{pole60} and \texttt{plik}+\texttt{equa30} masks. We interpret the deviation from unity visible at $\ell>1500$ in Figure~\ref{fig:sim_power_ratio} as indicating a difference in the effective beam functions. The red dashed line  shows a fit of the form $\exp\left[A(\ell/1000)^B\right]$, with parameters $A=1.03\times10^{-4}$ and $B=6.01$.

To quantitatively assess whether a window function change with this shape couples to cosmological parameters, we applied a by-hand rescaling of the data power spectrum in our 217~GHz likelihood file for the \texttt{plik}+\texttt{pole60} mask (essentially modifying the beam to match that from the \texttt{plik}+\texttt{equa30} mask, where the preference for $A_L>1$ was reduced). We found that the beam correction was absorbed into a slight trade-off between the CIB and point source foreground amplitudes, with minimal ($<0.1\sigma$) impact on $A_L$ or cosmological parameters. The $2.5\%$ deviation around $\ell=2500$ in Figure~\ref{fig:sim_power_ratio} corresponds to a $<0.5\sigma$ shift in the power spectrum bandpowers relative to the statistical uncertainty. Note that the 217~GHz spectra at $\ell>2000$ are increasingly noise dominated and carry little weight in the cosmological fitting. While the effective Planck beam functions will differ somewhat for every mask, and should in principle be recomputed each time, this test demonstrates that this is not a primary contributor to the preference for $A_L>1$ from the sky near the ecliptic.

\begin{table*}
  \centering
  \caption{Comparison between $A_L$ and $\Omega_k$ Constraints for Selection of Mask and Frequency Combinations}
  \begin{tabular}{lllccccc}
\hline
TT Spectra&TT Masking&Low-$\ell$ Data&$A_L$&$N\sigma$, $A_L>1$&$N\sigma$, $A_L>1$&$100\times\Omega_k$&$N\sigma$, $\Omega_k<0$\\
&&&(Mean, SD)&(Gaussian\footnote{Modeling posterior as Gaussian})&(Tail\footnote[2]{From fraction of MCMC chain points in $A_L\leq1$ tail (see text)})&(Mean, 68/95\%)&(Tail)\\
\hline
\hline
All&\texttt{plik}&\texttt{lowl}+\texttt{lowE}&$1.240\pm0.094$&$2.5\sigma$&$2.7\sigma$&$-5.5^{+2.3,+4.1}_{-2.3,-5.4}$&$3.0\sigma$\\
All&\texttt{plik}+\texttt{equa30}&\texttt{lowl}+\texttt{lowE}&$1.138\pm0.134$&$1.0\sigma$&$1.0\sigma$&$-3.5^{+2.5,+4.2}_{-2.5,-6.6}$&$1.5\sigma$\\
All&\texttt{plik}+\texttt{pole60}&\texttt{lowl}+\texttt{lowE}&$1.417\pm0.144$&$2.9\sigma$&$3.3\sigma$&$-11.0^{+4.6,+7.6}_{-4.5,-11.7}$&$3.5\sigma$\\
\hline
217&\texttt{plik}&\texttt{lowl}+\texttt{lowE}&$1.362\pm0.147$&$2.5\sigma$&$2.6\sigma$&$-8.6^{+4.2,+6.9}_{-4.2,-10.7}$&$2.9\sigma$\\
217&\texttt{plik}+\texttt{gal60\_857}&$$\texttt{lowl}+\texttt{lowE}&$1.182\pm0.151$&$1.2\sigma$&$1.2\sigma$&$-4.7^{+3.1,+5.0}_{-3.1,-7.8}$&$1.8\sigma$\\
\hline
100+143&\texttt{plik}&\texttt{lowl}+\texttt{lowE}&$1.198\pm0.109$&$1.8\sigma$&$1.9\sigma$&$-4.6^{+2.5,+4.2}_{-2.5,-6.0}$&$2.3\sigma$\\
100+143&\texttt{plik}&\texttt{lowE}&$1.144\pm0.112$&$1.3\sigma$&$1.3\sigma$&$-3.9^{+2.6,+4.5}_{-2.6,-6.2}$&$1.7\sigma$\\
\hline
\end{tabular}
\label{table:Ok_vs_AL}
\end{table*}

While the simple test described here is informative, it is far from an exhaustive check of Planck window function modeling or variability of Planck beam properties with sky area. For example, we have followed the Planck Collaboration in assuming that, for the power spectrum, the beam response can be adequately compressed to a 1D function (i.e., of $\ell$ alone). Testing this assumption would require analysis of lower-level Planck data products beyond the scope of the present work.

\section{Comparing $\Omega_k$ and $A_L$}
\label{sec:Ok}

We have focused on $A_L$ in the fits presented in this work. As mentioned in Section~\ref{sec:intro}, varying $\Omega_k$ while fixing $A_L=1$ improves the fit to the Planck TT data over \lcdm\ in a similar way to raising $A_L$. We emphasize again that the preference for $A_L>1$, for $\Omega_k<0$, and for different values of some \lcdm\ parameters with multipole range, are not independent issues but best viewed as three symptoms of the same underlying effect or effects in the Planck TT power spectrum.

In general, changing $A_L$ and changing $\Omega_k$ have distinct cosmological effects, with $\Omega_k$ impacting the geometry and expansion history of the universe and therefore quantities such as the angular diameter distance. Considering the Planck TT spectrum alone, however, without the addition of low-redshift cosmology measurements, varying $A_L$ and varying $\Omega_k$ produce similar deviations from \lcdm\ \citep[see, e.g., Figure~5 of][]{nguyen/huterer/wen:2023}. The structure of the acoustic peaks in the TT spectrum, in particular their spacing, is measured with high precision by Planck. For modest shifts in $\Omega_k$, this angular peak structure can be adequately maintained through shifts in other parameters impacting the expansion history (e.g., $H_0$ and $\Omega_m$). A second effect of changing $\Omega_k$ is to modify the post-recombination growth rate of structure, which for the purposes of the TT spectrum primarily impacts the lensing effect (i.e., the smoothing of the peaks). When $\Omega_k$ and the \lcdm\ parameters are varied simultaneously, there is sufficient freedom to alter the cosmic growth history and produce an enhancement of the lensing effect, mimicking the effect of $A_L>1$, while still maintaining a good fit for the acoustic peak spacing and larger angular scales where the lensing impact is minimal. A negative $\Omega_k$ also leads to a reduction in the late-time ISW effect, producing a slightly better fit to the $\ell<30$ TT data than when varying $A_L$ (see Section~\ref{sec:lowl} and Figure~\ref{fig:lowl}).

Table~\ref{table:Ok_vs_AL} shows $A_L$ and $\Omega_k$ constraints for a selection of mask and frequency combinations discussed in earlier Sections. The preference for $\Omega_k<0$ is always somewhat stronger than that for $A_L>1$, consistent with the fits performed by the Planck Collaboration \citep[e.g., Section~7.3 of ][]{planck2016-l06}. Overall, however, the $\Omega_k$ results show the same trends with frequency and mask selection as for $A_L$, consistent with the argument above that, when fitting TT data alone, the dominant effect of varying $\Omega_k$ is to change the strength of the lensing.

A complication when interpreting the $\Omega_k$ fits is that the 1D posterior distributions for $\Omega_k$ (marginalized over other cosmological and nuisance parameters) are more strongly non-Gaussian than for $A_L$ \citep[see Figure 19 of][]{tristram/etal:2024}. For this reason we quote the mean plus 68\% and 95\% intervals rather than just the mean and standard deviation (SD). One metric for the strength of preference for $\Omega_k<0$ is the fraction of chain points at $\Omega_k\geq0$ \citep[Section~7.3 of][]{planck2016-l06}. This can be translated into an equivalent number of `$\sigma$' deviations by asking how far one has to go in the tail of a Gaussian to find that same probability density. We report this value for both $\Omega_k$ and $A_L$ (using the fraction of points at $A_L\leq1$) in Table~\ref{table:Ok_vs_AL}. For $A_L$ we additionally provide the preference for $A_L>1$ simply assuming a Gaussian posterior, as used in the earlier Sections.

The results shown in Table~\ref{table:Ok_vs_AL} include only Planck temperature and (at low multipoles) polarization spectra. As mentioned in Section~\ref{sec:intro}, we do not interpret the preference for $\Omega_k<0$ from these data as evidence for physical curvature because many low redshift measurements (distance ladder, galaxy clustering, supernovae, etc.) are incompatible with the expansion and growth history predicted in these fits.

\section{Conclusions}
\label{sec:conclusions}

\subsection{Summary of Results}

We have reanalyzed Planck 2018 temperature maps, focusing on the phenomenological lensing amplitude parameter, $A_L$. As a point of comparison for the values quoted below, our $A_L$ constraint from Planck 2018 TT+\texttt{lowl}/\texttt{lowE} data, using the original \texttt{plik} masks, is $A_L=1.240\pm0.094$ (posterior mean and standard deviation), a $2.5\sigma$ preference for $A_L>1$. Our main findings are as follows:
\setlist[enumerate]{label={(\arabic*)}}
\begin{enumerate}
    \item The $A_L$ constraints depend on ecliptic latitude (Section~\ref{sec:ALecliptic}). Multiplying the \texttt{plik} masks by a mask admitting the half of the sky closest to the ecliptic poles yields $A_L=1.14\pm0.13$ ($1.0\sigma$). Using the complementary mask, admitting only the sky near the ecliptic, yields $A_L=1.41\pm0.14$ ($2.9\sigma$). The 217~GHz data shows the strongest single-spectrum dependence on the ecliptic masking, with $A_L=1.22\pm0.19$ near the ecliptic poles, $A_L=1.63\pm0.22$ near the ecliptic, and $1.36\pm0.15$ with the original \texttt{plik} mask.
    
    \item There is a significant excess of measured power in the 217~GHz data near the ecliptic (Section~\ref{sec:ecl_power}). From visual inspection of Planck and HI maps, we observed a tendency for brighter compact residual Galactic features admitted by the Planck masks to lie at low ecliptic latitudes. We therefore modified the \texttt{plik} masks to target these structures, using thresholds in 857~GHz intensity. For our \texttt{gal60\_857} mask, which reduces the admitted sky fraction from 0.50 to 0.42, the $A_L$ constraint shifts from $1.36\pm0.15$ to $1.18\pm0.15$. Based on simulations, this downward shift is incompatible with purely statistical fluctuations at the $3.1-3.5\sigma$ level. More stringent masking leads to larger uncertainties, however the mean 217-only $A_L$ values remain in the range $1.15-1.2$, in good agreement with the lower frequencies.
    
    \item The multifrequency $A_L$ constraints, with the 100 and 143~GHz data included, are less impacted by additional masking at 217~GHz (Section~\ref{sec:multifreq}). For the \texttt{gal60\_857} mask, the preference for $A_L>1$ remains at $2.5\sigma$. With even more stringent 217~GHz masking, using the \texttt{gal40\_857} mask, we find $1.21\pm0.10$ ($2.1\sigma$). Removing the 217~GHz data altogether yields $A_L=1.198\pm0.109$ ($1.8\sigma$).

    \item There appear to be at least two separate mechanisms contributing to the shifts in $A_L$ with masking at 217~GHz (Section~\ref{sec:eclplusdust}). Applying additional 217~GHz dust masking does not remove the preference for $A_L>1$ from low ecliptic latitudes. For example, adding the \texttt{gal60\_857} mask leads to a shift from $1.63\pm0.22$ to $1.52\pm0.23$, and the more stringent \texttt{gal40\_857} mask yields $1.67\pm0.32$.

    \item The $\ell<30$ TT spectrum has a moderate impact on the $A_L$ constraints, despite having no direct sensitivity to the lensing effect (Section~\ref{sec:lowl}). Removing the $\ell<30$ TT data reduces the preference for $A_L>1$ by $0.4\sigma$ for the multifrequency TT fit. Using WMAP instead of Planck for these multipoles reduces the preference for $A_L>1$ by $0.2\sigma$.

    \item We tested whether the lensing anomaly could be connected to residual CO line emission in the Planck maps, or dependence of the Planck effective beam function on ecliptic latitude, but found no evidence that these issues made any significant contribution ($<0.2\sigma$ in $A_L$, Section~\ref{sec:additional}).

    \item We investigated the effect of varying $\Omega_k$ instead of $A_L$ in fits to the Planck TT spectrum (Section~\ref{sec:Ok}). The $\Omega_k$ results show the same general trends with frequency and masking as $A_L$, although the preference for $\Omega_k<0$ over \lcdm\ is consistently larger than for $A_L>1$. This is partly due to a reduction of the late-time ISW effect in the $\Omega_k$ fits. 
    
    \item We found that the difference-of-covariance method \citep{gratton/challinor:2020} used by the Planck Collaboration to assess significance of shifts in cosmological constraints from nested data subsets is not valid for $A_L$ in some cases (Section~\ref{sec:ALshift217}). The posterior uncertainties  do not monotonically decrease when more data (e.g., more sky area) are added, contrary to simple Fisher information assumptions. This result motivates additional simulation-based checks in future analyses. 
    
\end{enumerate}

\subsection{Implications for the Origin of the Lensing Anomaly}

We set out to shed light on the origin of the $A_L$ anomaly in the Planck TT data, specifically, whether it could be attributed primarily to a genuine cosmological effect in the CMB (indicating some deficiency of the \lcdm\ model), a systematic error, or a statistical fluctuation. We have found that several separate effects contribute to the overall $2.5\sigma$ preference for $A_L>1$ from the TT spectrum. 

Removing the $\ell<30$ TT data reduces the anomaly to $2.1\sigma$ ($A_L=1.205\pm0.096$). An increase in $A_L$ is accompanied by shifts in correlated \lcdm\ parameters, including a decrease in $A_s$ and an increase in $n_s$. This reduces the model power at low $\ell$, better matching the WMAP and Planck measurements, which are generally low of the \lcdm\ model fit in this range, particularly for the `dip' at $20\leq\ell\leq22$ (Figure~\ref{fig:lowl}). Whether this mismatch with \lcdm\ is a genuine sign of a problem with the model or just a statistical fluke has been debated extensively in the literature, often alongside other seemingly anomalous features of the large-scale temperature modes \citep[e.g., hemispherical asymmetry, alignment of modes at different multipoles, etc.; see][for a review]{schwarz/etal:2016}. Our current analysis does not weigh in on this question, and we have no evidence to support an exclusion of the low multipole data. For the purposes of $A_L$, it is useful to separate the impact of the Planck $\ell<30$ data because these scales were analyzed separately from the higher multipoles (using a pixel-based likelihood, using a very different map-based foreground cleaning approach, and using a larger fraction of the sky). Many foreground-related systematic residuals, for example, would be effectively independent at $\ell<30$ compared to $\ell>30$.

We found in Section~\ref{sec:lowl} that using WMAP instead of Planck for the low-multipole temperature likelihood reduces the preference for $A_L>1$ by $0.2\sigma$ ($A_L=1.223\pm0.095$). This difference, being half the shift from dropping $\ell<30$ TT altogether, is perhaps surprising, given that both experiments are highly cosmic variance limited on these scales. Separately from the origins of the lensing anomaly, it motivates revisiting the masking and foreground treatment for the large-scale pixel-based likelihoods.

We found that a portion of the remaining $2.1\sigma$ preference for $A_L>1$ is contributed by the 217~GHz channel, and that this is not robust to sky masking choices. First, there is a preference for higher $A_L$ values from the portion of the sky closest to the ecliptic. This preference is strongest at 217~GHz, and weaker or absent at 100 and 143~GHz (Figure~\ref{fig:ecliptic_AL_1d}). While we have no direct evidence for a systematic origin for the ecliptic latitude dependence, the connection with the Planck scanning strategy is suggestive of some non-random effect. Second, separately from the ecliptic latitude dependence, we found that additional dust masking significantly shifts the 217~GHz $A_L$ constraint downwards, bringing it into better agreement with the other frequencies (with a mean in the range $1.15-1.2$, for fits including the $\ell<30$ TT data). This shift is too large to be explained solely by statistical fluctuations, as estimated using simulations where the true $A_L$ is unity. In light of these results, a reasonable conservative approach could be to remove the 217~GHz TT data from the $A_L$ analysis altogether. Doing so yields $A_L=1.198\pm0.109$ ($1.8\sigma$) with the $\ell<30$ data included, or $A_L=1.144\pm0.112$ ($1.3\sigma$) using only the high multipole TT likelihood.

This remaining $1.3\sigma$ preference for $A_L>1$ from the high multipole 100 and 143~GHz data appears robust to choice of sky mask and other tests we have performed in this work. Clearly, a statistical fluctuation at this level would not be surprising, and we are not able to distinguish this possibility from a cosmological or systematic error origin.

\subsection{Broader Implications}

What bearing does this work have on the prevailing cosmological tensions involving $H_0$ or $S_8$? As stated in Section~\ref{sec:intro}, the $H_0$ tension exists independently of Planck, and therefore could not be resolved through any change in the treatment of the Planck data. Recent ACT CMB measurements also support the higher values of $S_8$ preferred by Planck \citep{aiola/etal:2020,madhavacheril/etal:2024}. There is, however, a question as to how whatever underlying causes contribute to the preference to $A_L>1$ may impact fits performed to constrain alternative cosmological models. Marginalizing over $A_L$ can provide a check of this, although having $A_L$ only modify the theory model being compared to the Planck TT data (and not the theory spectra being compared to polarization or lensing data, or other CMB data sets) may provide a useful alternative test. One could also test the effect of excluding the Planck 217~GHz data from the expanded model fit. Within \lcdm, the improved angular resolution and additional foreground information provided by 217~GHz has little impact on parameter constraints, and removing the $143\times217$ and 217~GHz spectra leads to shifts around the $0.2\sigma$ level or smaller, even considering only TT data (Figure~78 of PL18). This may not be the case for all model extensions, particularly those with extra degrees of freedom impacting multipoles $\ell\geq1500$.

Finally, one aspect of the ecliptic latitude dependence worth highlighting is the relation to regions observed by high-resolution ground-based CMB telescopes. The SPT fields are, naturally, close to the equatorial south pole, at declinations below $-42\degree$ \citep[Figure~1 of][]{dutcher/etal:2021}, and outside the low ecliptic latitude region that shows the strongest preference for $A_L>1$ in the Planck TT data. In contrast, there are several deep ACT fields along the celestial equator, which are close to the ecliptic plane \citep[Figure~1 of][]{aiola/etal:2020}. This may enable more powerful checks of the Planck preference for $A_L>1$ at low ecliptic latitudes, for instance using upcoming 230~GHz Advanced ACTPol data \citep{henderson/etal:2016}.\\

GEA thanks Eric Hivon and Karim Benabed for clarifications relating to the Planck analysis and public data products. This work was supported in part by NASA ROSES grants 80NSSC22K0408 and 80NSSC23K0475. This work was based on observations obtained with Planck (http://www.esa.int/Planck), an ESA science mission with instruments and contributions directly funded by ESA Member States, NASA, and Canada. We acknowledge the use of the Legacy Archive for Microwave Background Data Analysis (LAMBDA), part of the High Energy Astrophysics Science Archive Center (HEASARC). HEASARC/LAMBDA is a service of the Astrophysics Science Division at the NASA Goddard Space Flight Center. Some calculations for this work were carried out at the Advanced Research Computing at Hopkins (ARCH) core facility (rockfish.jhu.edu), which is supported by the National Science Foundation (NSF) grant number OAC 1920103.

\software{numpy \citep{harris/etal:2020}, scipy \citep{virtanen/etal:2020}, matplotlib \citep{hunter:2007}, astropy \citep{astropy:2013,astropy2018,astropy2022}, HEALPix \citep{gorski2005}, CAMB \citep{lewis/challinor/lasenby:2000,howlett/etal:2012}, CosmoMC \citep{lewis/bridle:2002}, GetDist \citep{getdist2019}}

\bibliographystyle{apj}

\begin{thebibliography}{}
\expandafter\ifx\csname natexlab\endcsname\relax\def\natexlab#1{#1}\fi

\bibitem[{{Addison} {et~al.}(2016){Addison}, {Huang}, {Watts}, {Bennett},
  {Halpern}, {Hinshaw}, \& {Weiland}}]{addison/etal:2016}
{Addison}, G.~E., {Huang}, Y., {Watts}, D.~J., {et~al.} 2016, \apj, 818, 132

\bibitem[{{Addison} {et~al.}(2018){Addison}, {Watts}, {Bennett}, {Halpern},
  {Hinshaw}, \& {Weiland}}]{addison/etal:2018}
{Addison}, G.~E., {Watts}, D.~J., {Bennett}, C.~L., {et~al.} 2018, \apj, 853,
  119

\bibitem[{{Aiola} {et~al.}(2020){Aiola}, {Calabrese}, {Maurin}, {Naess},
  {Schmitt}, {Abitbol}, {Addison}, {Ade}, {Alonso}, {Amiri}, {Amodeo},
  {Angile}, {Austermann}, {Baildon}, {Battaglia}, {Beall}, {Bean}, {Becker},
  {Bond}, {Bruno}, {Calafut}, {Campusano}, {Carrero}, {Chesmore}, {Cho},
  {Choi}, {Clark}, {Cothard}, {Crichton}, {Crowley}, {Darwish}, {Datta},
  {Denison}, {Devlin}, {Duell}, {Duff}, {Duivenvoorden}, {Dunkley},
  {D{\"u}nner}, {Essinger-Hileman}, {Fankhanel}, {Ferraro}, {Fox}, {Fuzia},
  {Gallardo}, {Gluscevic}, {Golec}, {Grace}, {Gralla}, {Guan}, {Hall},
  {Halpern}, {Han}, {Hargrave}, {Hasselfield}, {Helton}, {Henderson},
  {Hensley}, {Hill}, {Hilton}, {Hilton}, {Hincks}, {Hlo{\v{z}}ek}, {Ho},
  {Hubmayr}, {Huffenberger}, {Hughes}, {Infante}, {Irwin}, {Jackson}, {Klein},
  {Knowles}, {Koopman}, {Kosowsky}, {Lakey}, {Li}, {Li}, {Li}, {Lokken},
  {Louis}, {Lungu}, {MacInnis}, {Madhavacheril}, {Maldonado}, {Mallaby-Kay},
  {Marsden}, {McMahon}, {Menanteau}, {Moodley}, {Morton}, {Namikawa}, {Nati},
  {Newburgh}, {Nibarger}, {Nicola}, {Niemack}, {Nolta}, {Orlowski-Sherer},
  {Page}, {Pappas}, {Partridge}, {Phakathi}, {Pisano}, {Prince}, {Puddu}, {Qu},
  {Rivera}, {Robertson}, {Rojas}, {Salatino}, {Schaan}, {Schillaci}, {Sehgal},
  {Sherwin}, {Sierra}, {Sievers}, {Sifon}, {Sikhosana}, {Simon}, {Spergel},
  {Staggs}, {Stevens}, {Storer}, {Sunder}, {Switzer}, {Thorne}, {Thornton},
  {Trac}, {Treu}, {Tucker}, {Vale}, {Van Engelen}, {Van Lanen}, {Vavagiakis},
  {Wagoner}, {Wang}, {Ward}, {Wollack}, {Xu}, {Zago}, \&
  {Zhu}}]{aiola/etal:2020}
{Aiola}, S., {Calabrese}, E., {Maurin}, L., {et~al.} 2020, \jcap, 2020, 047

\bibitem[{{Alam} {et~al.}(2021){Alam}, {Aubert}, {Avila}, {Balland},
  {Bautista}, {Bershady}, {Bizyaev}, {Blanton}, {Bolton}, {Bovy}, {Brinkmann},
  {Brownstein}, {Burtin}, {Chabanier}, {Chapman}, {Choi}, {Chuang}, {Comparat},
  {Cousinou}, {Cuceu}, {Dawson}, {de la Torre}, {de Mattia}, {Agathe}, {des
  Bourboux}, {Escoffier}, {Etourneau}, {Farr}, {Font-Ribera}, {Frinchaboy},
  {Fromenteau}, {Gil-Mar{\'\i}n}, {Le Goff}, {Gonzalez-Morales},
  {Gonzalez-Perez}, {Grabowski}, {Guy}, {Hawken}, {Hou}, {Kong}, {Parker},
  {Klaene}, {Kneib}, {Lin}, {Long}, {Lyke}, {de la Macorra}, {Martini},
  {Masters}, {Mohammad}, {Moon}, {Mueller}, {Mu{\~n}oz-Guti{\'e}rrez}, {Myers},
  {Nadathur}, {Neveux}, {Newman}, {Noterdaeme}, {Oravetz}, {Oravetz},
  {Palanque-Delabrouille}, {Pan}, {Paviot}, {Percival}, {P{\'e}rez-R{\`a}fols},
  {Petitjean}, {Pieri}, {Prakash}, {Raichoor}, {Ravoux}, {Rezaie}, {Rich},
  {Ross}, {Rossi}, {Ruggeri}, {Ruhlmann-Kleider}, {S{\'a}nchez}, {S{\'a}nchez},
  {S{\'a}nchez-Gallego}, {Sayres}, {Schneider}, {Seo}, {Shafieloo}, {Slosar},
  {Smith}, {Stermer}, {Tamone}, {Tinker}, {Tojeiro}, {Vargas-Maga{\~n}a},
  {Variu}, {Wang}, {Weaver}, {Weijmans}, {Y{\`e}che}, {Zarrouk}, {Zhao},
  {Zhao}, \& {Zheng}}]{alam/etal:2021}
{Alam}, S., {Aubert}, M., {Avila}, S., {et~al.} 2021, \prd, 103, 083533

\bibitem[{{Alonso} {et~al.}(2019){Alonso}, {Sanchez}, {Slosar}, \& {LSST Dark
  Energy Science Collaboration}}]{alonso/etal:2019}
{Alonso}, D., {Sanchez}, J., {Slosar}, A., \& {LSST Dark Energy Science
  Collaboration}. 2019, \mnras, 484, 4127

\bibitem[{{Astropy Collaboration} {et~al.}(2013){Astropy Collaboration},
  {Robitaille}, {Tollerud}, {Greenfield}, {Droettboom}, {Bray}, {Aldcroft},
  {Davis}, {Ginsburg}, {Price-Whelan}, {Kerzendorf}, {Conley}, {Crighton},
  {Barbary}, {Muna}, {Ferguson}, {Grollier}, {Parikh}, {Nair}, {Unther},
  {Deil}, {Woillez}, {Conseil}, {Kramer}, {Turner}, {Singer}, {Fox}, {Weaver},
  {Zabalza}, {Edwards}, {Azalee Bostroem}, {Burke}, {Casey}, {Crawford},
  {Dencheva}, {Ely}, {Jenness}, {Labrie}, {Lim}, {Pierfederici}, {Pontzen},
  {Ptak}, {Refsdal}, {Servillat}, \& {Streicher}}]{astropy:2013}
{Astropy Collaboration}, {Robitaille}, T.~P., {Tollerud}, E.~J., {et~al.} 2013,
  \aap, 558, A33

\bibitem[{{Astropy Collaboration} {et~al.}(2018){Astropy Collaboration},
  {Price-Whelan}, {Sip{\H{o}}cz}, {G{\"u}nther}, {Lim}, {Crawford}, {Conseil},
  {Shupe}, {Craig}, {Dencheva}, {Ginsburg}, {VanderPlas}, {Bradley},
  {P{\'e}rez-Su{\'a}rez}, {de Val-Borro}, {Aldcroft}, {Cruz}, {Robitaille},
  {Tollerud}, {Ardelean}, {Babej}, {Bach}, {Bachetti}, {Bakanov}, {Bamford},
  {Barentsen}, {Barmby}, {Baumbach}, {Berry}, {Biscani}, {Boquien}, {Bostroem},
  {Bouma}, {Brammer}, {Bray}, {Breytenbach}, {Buddelmeijer}, {Burke},
  {Calderone}, {Cano Rodr{\'\i}guez}, {Cara}, {Cardoso}, {Cheedella}, {Copin},
  {Corrales}, {Crichton}, {D'Avella}, {Deil}, {Depagne}, {Dietrich}, {Donath},
  {Droettboom}, {Earl}, {Erben}, {Fabbro}, {Ferreira}, {Finethy}, {Fox},
  {Garrison}, {Gibbons}, {Goldstein}, {Gommers}, {Greco}, {Greenfield},
  {Groener}, {Grollier}, {Hagen}, {Hirst}, {Homeier}, {Horton}, {Hosseinzadeh},
  {Hu}, {Hunkeler}, {Ivezi{\'c}}, {Jain}, {Jenness}, {Kanarek}, {Kendrew},
  {Kern}, {Kerzendorf}, {Khvalko}, {King}, {Kirkby}, {Kulkarni}, {Kumar},
  {Lee}, {Lenz}, {Littlefair}, {Ma}, {Macleod}, {Mastropietro}, {McCully},
  {Montagnac}, {Morris}, {Mueller}, {Mumford}, {Muna}, {Murphy}, {Nelson},
  {Nguyen}, {Ninan}, {N{\"o}the}, {Ogaz}, {Oh}, {Parejko}, {Parley}, {Pascual},
  {Patil}, {Patil}, {Plunkett}, {Prochaska}, {Rastogi}, {Reddy Janga},
  {Sabater}, {Sakurikar}, {Seifert}, {Sherbert}, {Sherwood-Taylor}, {Shih},
  {Sick}, {Silbiger}, {Singanamalla}, {Singer}, {Sladen}, {Sooley},
  {Sornarajah}, {Streicher}, {Teuben}, {Thomas}, {Tremblay}, {Turner},
  {Terr{\'o}n}, {van Kerkwijk}, {de la Vega}, {Watkins}, {Weaver}, {Whitmore},
  {Woillez}, {Zabalza}, \& {Astropy Contributors}}]{astropy2018}
{Astropy Collaboration}, {Price-Whelan}, A.~M., {Sip{\H{o}}cz}, B.~M., {et~al.}
  2018, \aj, 156, 123

\bibitem[{{Astropy Collaboration} {et~al.}(2022){Astropy Collaboration},
  {Price-Whelan}, {Lim}, {Earl}, {Starkman}, {Bradley}, {Shupe}, {Patil},
  {Corrales}, {Brasseur}, {N{\"o}the}, {Donath}, {Tollerud}, {Morris},
  {Ginsburg}, {Vaher}, {Weaver}, {Tocknell}, {Jamieson}, {van Kerkwijk},
  {Robitaille}, {Merry}, {Bachetti}, {G{\"u}nther}, {Aldcroft},
  {Alvarado-Montes}, {Archibald}, {B{\'o}di}, {Bapat}, {Barentsen},
  {Baz{\'a}n}, {Biswas}, {Boquien}, {Burke}, {Cara}, {Cara}, {Conroy},
  {Conseil}, {Craig}, {Cross}, {Cruz}, {D'Eugenio}, {Dencheva}, {Devillepoix},
  {Dietrich}, {Eigenbrot}, {Erben}, {Ferreira}, {Foreman-Mackey}, {Fox},
  {Freij}, {Garg}, {Geda}, {Glattly}, {Gondhalekar}, {Gordon}, {Grant},
  {Greenfield}, {Groener}, {Guest}, {Gurovich}, {Handberg}, {Hart},
  {Hatfield-Dodds}, {Homeier}, {Hosseinzadeh}, {Jenness}, {Jones}, {Joseph},
  {Kalmbach}, {Karamehmetoglu}, {Ka{\l}uszy{\'n}ski}, {Kelley}, {Kern},
  {Kerzendorf}, {Koch}, {Kulumani}, {Lee}, {Ly}, {Ma}, {MacBride}, {Maljaars},
  {Muna}, {Murphy}, {Norman}, {O'Steen}, {Oman}, {Pacifici}, {Pascual},
  {Pascual-Granado}, {Patil}, {Perren}, {Pickering}, {Rastogi}, {Roulston},
  {Ryan}, {Rykoff}, {Sabater}, {Sakurikar}, {Salgado}, {Sanghi}, {Saunders},
  {Savchenko}, {Schwardt}, {Seifert-Eckert}, {Shih}, {Jain}, {Shukla}, {Sick},
  {Simpson}, {Singanamalla}, {Singer}, {Singhal}, {Sinha}, {Sip{\H{o}}cz},
  {Spitler}, {Stansby}, {Streicher}, {{\v{S}}umak}, {Swinbank}, {Taranu},
  {Tewary}, {Tremblay}, {de Val-Borro}, {Van Kooten}, {Vasovi{\'c}}, {Verma},
  {de Miranda Cardoso}, {Williams}, {Wilson}, {Winkel}, {Wood-Vasey}, {Xue},
  {Yoachim}, {Zhang}, {Zonca}, \& {Astropy Project Contributors}}]{astropy2022}
{Astropy Collaboration}, {Price-Whelan}, A.~M., {Lim}, P.~L., {et~al.} 2022,
  \apj, 935, 167

\bibitem[{{Aubourg} {et~al.}(2015){Aubourg}, {Bailey}, {Bautista}, {Beutler},
  {Bhardwaj}, {Bizyaev}, {Blanton}, {Blomqvist}, {Bolton}, {Bovy},
  {Brewington}, {Brinkmann}, {Brownstein}, {Burden}, {Busca}, {Carithers},
  {Chuang}, {Comparat}, {Croft}, {Cuesta}, {Dawson}, {Delubac}, {Eisenstein},
  {Font-Ribera}, {Ge}, {Le Goff}, {Gontcho}, {Gott}, {Gunn}, {Guo}, {Guy},
  {Hamilton}, {Ho}, {Honscheid}, {Howlett}, {Kirkby}, {Kitaura}, {Kneib},
  {Lee}, {Long}, {Lupton}, {Maga{\~n}a}, {Malanushenko}, {Malanushenko},
  {Manera}, {Maraston}, {Margala}, {McBride}, {Miralda-Escud{\'e}}, {Myers},
  {Nichol}, {Noterdaeme}, {Nuza}, {Olmstead}, {Oravetz}, {P{\^a}ris},
  {Padmanabhan}, {Palanque-Delabrouille}, {Pan}, {Pellejero-Ibanez},
  {Percival}, {Petitjean}, {Pieri}, {Prada}, {Reid}, {Rich}, {Roe}, {Ross},
  {Ross}, {Rossi}, {Rubi{\~n}o-Mart{\'{\i}}n}, {S{\'a}nchez}, {Samushia},
  {G{\'e}nova-Santos}, {Sc{\'o}ccola}, {Schlegel}, {Schneider}, {Seo},
  {Sheldon}, {Simmons}, {Skibba}, {Slosar}, {Strauss}, {Thomas}, {Tinker},
  {Tojeiro}, {Vazquez}, {Viel}, {Wake}, {Weaver}, {Weinberg}, {Wood-Vasey},
  {Y{\`e}che}, {Zehavi}, {Zhao}, \& {BOSS Collaboration}}]{aubourg/etal:2015}
{Aubourg}, {\'E}., {Bailey}, S., {Bautista}, J.~E., {et~al.} 2015, \prd, 92,
  123516

\bibitem[{{Bennett} {et~al.}(2011){Bennett}, {Hill}, {Hinshaw}, {Larson},
  {Smith}, {Dunkley}, {Gold}, {Halpern}, {Jarosik}, {Kogut}, {Komatsu},
  {Limon}, {Meyer}, {Nolta}, {Odegard}, {Page}, {Spergel}, {Tucker}, {Weiland},
  {Wollack}, \& {Wright}}]{bennett/etal:2011}
{Bennett}, C.~L., {Hill}, R.~S., {Hinshaw}, G., {et~al.} 2011, \apjs, 192, 17

\bibitem[{{Bennett} {et~al.}(2013){Bennett}, {Larson}, {Weiland}, {Jarosik},
  {Hinshaw}, {Odegard}, {Smith}, {Hill}, {Gold}, {Halpern}, {Komatsu}, {Nolta},
  {Page}, {Spergel}, {Wollack}, {Dunkley}, {Kogut}, {Limon}, {Meyer}, {Tucker},
  \& {Wright}}]{bennett/etal:2013}
{Bennett}, C.~L., {Larson}, D., {Weiland}, J.~L., {et~al.} 2013, \apjs, 208, 20

\bibitem[{{Calabrese} {et~al.}(2008){Calabrese}, {Slosar}, {Melchiorri},
  {Smoot}, \& {Zahn}}]{calabrese/etal:2008}
{Calabrese}, E., {Slosar}, A., {Melchiorri}, A., {Smoot}, G.~F., \& {Zahn}, O.
  2008, \prd, 77, 123531

\bibitem[{{Carron} {et~al.}(2022){Carron}, {Mirmelstein}, \&
  {Lewis}}]{carron/mirmelstein/lewis:2022}
{Carron}, J., {Mirmelstein}, M., \& {Lewis}, A. 2022, \jcap, 2022, 039

\bibitem[{{Choi} {et~al.}(2020){Choi}, {Hasselfield}, {Ho}, {Koopman}, {Lungu},
  {Abitbol}, {Addison}, {Ade}, {Aiola}, {Alonso}, {Amiri}, {Amodeo}, {Angile},
  {Austermann}, {Baildon}, {Battaglia}, {Beall}, {Bean}, {Becker}, {Bond},
  {Bruno}, {Calabrese}, {Calafut}, {Campusano}, {Carrero}, {Chesmore}, {Cho},
  {Clark}, {Cothard}, {Crichton}, {Crowley}, {Darwish}, {Datta}, {Denison},
  {Devlin}, {Duell}, {Duff}, {Duivenvoorden}, {Dunkley}, {D{\"u}nner},
  {Essinger-Hileman}, {Fankhanel}, {Ferraro}, {Fox}, {Fuzia}, {Gallardo},
  {Gluscevic}, {Golec}, {Grace}, {Gralla}, {Guan}, {Hall}, {Halpern}, {Han},
  {Hargrave}, {Henderson}, {Hensley}, {Hill}, {Hilton}, {Hilton}, {Hincks},
  {Hlo{\v{z}}ek}, {Hubmayr}, {Huffenberger}, {Hughes}, {Infante}, {Irwin},
  {Jackson}, {Klein}, {Knowles}, {Kosowsky}, {Lakey}, {Li}, {Li}, {Li},
  {Lokken}, {Louis}, {MacInnis}, {Madhavacheril}, {Maldonado}, {Mallaby-Kay},
  {Marsden}, {Maurin}, {McMahon}, {Menanteau}, {Moodley}, {Morton}, {Naess},
  {Namikawa}, {Nati}, {Newburgh}, {Nibarger}, {Nicola}, {Niemack}, {Nolta},
  {Orlowski-Sherer}, {Page}, {Pappas}, {Partridge}, {Phakathi}, {Prince},
  {Puddu}, {Qu}, {Rivera}, {Robertson}, {Rojas}, {Salatino}, {Schaan},
  {Schillaci}, {Schmitt}, {Sehgal}, {Sherwin}, {Sierra}, {Sievers}, {Sifon},
  {Sikhosana}, {Simon}, {Spergel}, {Staggs}, {Stevens}, {Storer}, {Sunder},
  {Switzer}, {Thorne}, {Thornton}, {Trac}, {Treu}, {Tucker}, {Vale}, {Van
  Engelen}, {Van Lanen}, {Vavagiakis}, {Wagoner}, {Wang}, {Ward}, {Wollack},
  {Xu}, {Zago}, \& {Zhu}}]{choi/etal:2020}
{Choi}, S.~K., {Hasselfield}, M., {Ho}, S.-P.~P., {et~al.} 2020, \jcap, 2020,
  045

\bibitem[{{Dark Energy Survey and Kilo-Degree Survey Collaboration}
  {et~al.}(2023){Dark Energy Survey and Kilo-Degree Survey Collaboration},
  {Abbott}, {Aguena}, {Alarcon}, {Alves}, {Amon}, {Andrade-Oliveira}, {Asgari},
  {Avila}, {Bacon}, {Bechtol}, {Becker}, {Bernstein}, {Bertin}, {Bilicki},
  {Blazek}, {Bocquet}, {Brooks}, {Burger}, {Burke}, {Camacho}, {Campos},
  {Carnero Rosell}, {Carrasco Kind}, {Carretero}, {Castander}, {Cawthon},
  {Chang}, {Chen}, {Choi}, {Conselice}, {Cordero}, {Crocce}, {da Costa}, {da
  Silva Pereira}, {Dalal}, {Davis}, {de Jong}, {DeRose}, {Desai}, {Diehl},
  {Dodelson}, {Doel}, {Doux}, {Drlica-Wagner}, {Dvornik}, {Eckert}, {Eifler},
  {Elvin-Poole}, {Everett}, {Fang}, {Ferrero}, {Fert{\'e}}, {Flaugher},
  {Friedrich}, {Frieman}, {Garc{\'\i}a-Bellido}, {Gatti}, {Giannini}, {Giblin},
  {Gruen}, {Gruendl}, {Gutierrez}, {Harrison}, {Hartley}, {Herner}, {Heymans},
  {Hildebrandt}, {Hinton}, {Hoekstra}, {Hollowood}, {Honscheid}, {Huang},
  {Huff}, {Huterer}, {James}, {Jarvis}, {Jeffrey}, {Jeltema}, {Joachimi},
  {Joudaki}, {Kannawadi}, {Krause}, {Kuehn}, {Kuijken}, {Kuropatkin}, {Lahav},
  {Leget}, {Lemos}, {Li}, {Li}, {Liddle}, {Lima}, {Lin}, {Lin}, {MacCrann},
  {Mahony}, {Marshall}, {McCullough}, {Mena-Fern{\'a}ndez}, {Menanteau},
  {Miquel}, {Mohr}, {Muir}, {Myles}, {Napolitano}, {Navarro-Alsina}, {Ogando},
  {Palmese}, {Pandey}, {Park}, {Paterno}, {Peacock}, {Petravick}, {Pieres},
  {Plazas Malag{\'o}n}, {Porredon}, {Prat}, {Radovich}, {Raveri}, {Reischke},
  {Robertson}, {Rollins}, {Romer}, {Roodman}, {Rykoff}, {Samuroff},
  {S{\'a}nchez}, {Sanchez}, {Sanchez}, {Schneider}, {Secco}, {Sevilla-Noarbe},
  {Shan}, {Sheldon}, {Shin}, {Sif{\'o}n}, {Smith}, {Soares-Santos},
  {St{\"o}lzner}, {Suchyta}, {Swanson}, {Tarle}, {Thomas}, {To}, {Troxel},
  {Tr{\"o}ster}, {Tutusaus}, {van den Busch}, {Varga}, {Walker}, {Weaverdyck},
  {Wechsler}, {Weller}, {Wiseman}, {Wright}, {Yanny}, {Yin}, {Yoon}, {Zhang},
  \& {Zuntz}}]{des/kids:2023}
{Dark Energy Survey and Kilo-Degree Survey Collaboration}, {Abbott}, T.~M.~C.,
  {Aguena}, M., {et~al.} 2023, The Open Journal of Astrophysics, 6, 36

\bibitem[{{Das} {et~al.}(2011){Das}, {Sherwin}, {Aguirre}, {Appel}, {Bond},
  {Carvalho}, {Devlin}, {Dunkley}, {D{\"u}nner}, {Essinger-Hileman}, {Fowler},
  {Hajian}, {Halpern}, {Hasselfield}, {Hincks}, {Hlozek}, {Huffenberger},
  {Hughes}, {Irwin}, {Klein}, {Kosowsky}, {Lupton}, {Marriage}, {Marsden},
  {Menanteau}, {Moodley}, {Niemack}, {Nolta}, {Page}, {Parker}, {Reese},
  {Schmitt}, {Sehgal}, {Sievers}, {Spergel}, {Staggs}, {Swetz}, {Switzer},
  {Thornton}, {Visnjic}, \& {Wollack}}]{das/etal:2011}
{Das}, S., {Sherwin}, B.~D., {Aguirre}, P., {et~al.} 2011, \prl, 107, 021301

\bibitem[{{Di Valentino} {et~al.}(2020){Di Valentino}, {Melchiorri}, \&
  {Silk}}]{divalentino/melchiorri/silk:2020}
{Di Valentino}, E., {Melchiorri}, A., \& {Silk}, J. 2020, Nature Astronomy, 4,
  196

\bibitem[{{Dutcher} {et~al.}(2021){Dutcher}, {Balkenhol}, {Ade}, {Ahmed},
  {Anderes}, {Anderson}, {Archipley}, {Avva}, {Aylor}, {Barry}, {Basu Thakur},
  {Benabed}, {Bender}, {Benson}, {Bianchini}, {Bleem}, {Bouchet}, {Bryant},
  {Byrum}, {Carlstrom}, {Carter}, {Cecil}, {Chang}, {Chaubal}, {Chen}, {Cho},
  {Chou}, {Cliche}, {Crawford}, {Cukierman}, {Daley}, {de Haan}, {Denison},
  {Dibert}, {Ding}, {Dobbs}, {Everett}, {Feng}, {Ferguson}, {Foster}, {Fu},
  {Galli}, {Gambrel}, {Gardner}, {Goeckner-Wald}, {Gualtieri}, {Guns}, {Gupta},
  {Guyser}, {Halverson}, {Harke-Hosemann}, {Harrington}, {Henning}, {Hilton},
  {Hivon}, {Holder}, {Holzapfel}, {Hood}, {Howe}, {Huang}, {Irwin}, {Jeong},
  {Jonas}, {Jones}, {Khaire}, {Knox}, {Kofman}, {Korman}, {Kubik}, {Kuhlmann},
  {Kuo}, {Lee}, {Leitch}, {Lowitz}, {Lu}, {Meyer}, {Michalik}, {Millea},
  {Montgomery}, {Nadolski}, {Natoli}, {Nguyen}, {Noble}, {Novosad}, {Omori},
  {Padin}, {Pan}, {Paschos}, {Pearson}, {Posada}, {Prabhu}, {Quan},
  {Raghunathan}, {Rahlin}, {Reichardt}, {Riebel}, {Riedel}, {Rouble}, {Ruhl},
  {Sayre}, {Schiappucci}, {Shirokoff}, {Smecher}, {Sobrin}, {Stark}, {Stephen},
  {Story}, {Suzuki}, {Thompson}, {Thorne}, {Tucker}, {Umilta}, {Vale},
  {Vanderlinde}, {Vieira}, {Wang}, {Whitehorn}, {Wu}, {Yefremenko}, {Yoon},
  {Young}, \& {SPT-3G Collaboration}}]{dutcher/etal:2021}
{Dutcher}, D., {Balkenhol}, L., {Ade}, P.~A.~R., {et~al.} 2021, \prd, 104,
  022003

\bibitem[{{Efstathiou}(2004)}]{efstathiou:2004}
{Efstathiou}, G. 2004, \mnras, 349, 603

\bibitem[{{Efstathiou} \& {Gratton}(2021)}]{efstathiou/gratton:2021}
{Efstathiou}, G., \& {Gratton}, S. 2021, The Open Journal of Astrophysics, 4, 8

\bibitem[{{G{\'o}rski} {et~al.}(2005){G{\'o}rski}, {Hivon}, {Banday},
  {Wandelt}, {Hansen}, {Reinecke}, \& {Bartelmann}}]{gorski2005}
{G{\'o}rski}, K.~M., {Hivon}, E., {Banday}, A.~J., {et~al.} 2005, \apj, 622,
  759

\bibitem[{{Gratton} \& {Challinor}(2020)}]{gratton/challinor:2020}
{Gratton}, S., \& {Challinor}, A. 2020, \mnras, 499, 3410

\bibitem[{{Gruetjen} {et~al.}(2017){Gruetjen}, {Fergusson}, {Liguori}, \&
  {Shellard}}]{gruetjen/etal:2017}
{Gruetjen}, H.~F., {Fergusson}, J.~R., {Liguori}, M., \& {Shellard}, E.~P.~S.
  2017, \prd, 95, 043532

\bibitem[{Handley(2021)}]{handley:2021}
Handley, W. 2021, Phys. Rev. D, 103, L041301

\bibitem[{Harris {et~al.}(2020)Harris, Millman, van~der Walt, Gommers,
  Virtanen, Cournapeau, Wieser, Taylor, Berg, Smith, Kern, Picus, Hoyer, van
  Kerkwijk, Brett, Haldane, del R{\'{i}}o, Wiebe, Peterson,
  G{\'{e}}rard-Marchant, Sheppard, Reddy, Weckesser, Abbasi, Gohlke, \&
  Oliphant}]{harris/etal:2020}
Harris, C.~R., Millman, K.~J., van~der Walt, S.~J., {et~al.} 2020, Nature, 585,
  357

\bibitem[{{Henderson} {et~al.}(2016){Henderson}, {Allison}, {Austermann},
  {Baildon}, {Battaglia}, {Beall}, {Becker}, {De Bernardis}, {Bond},
  {Calabrese}, {Choi}, {Coughlin}, {Crowley}, {Datta}, {Devlin}, {Duff},
  {Dunkley}, {D{\"u}nner}, {van Engelen}, {Gallardo}, {Grace}, {Hasselfield},
  {Hills}, {Hilton}, {Hincks}, {Hlo?ek}, {Ho}, {Hubmayr}, {Huffenberger},
  {Hughes}, {Irwin}, {Koopman}, {Kosowsky}, {Li}, {McMahon}, {Munson}, {Nati},
  {Newburgh}, {Niemack}, {Niraula}, {Page}, {Pappas}, {Salatino}, {Schillaci},
  {Schmitt}, {Sehgal}, {Sherwin}, {Sievers}, {Simon}, {Spergel}, {Staggs},
  {Stevens}, {Thornton}, {Van Lanen}, {Vavagiakis}, {Ward}, \&
  {Wollack}}]{henderson/etal:2016}
{Henderson}, S.~W., {Allison}, R., {Austermann}, J., {et~al.} 2016, Journal of
  Low Temperature Physics, 184, 772

\bibitem[{{Henning} {et~al.}(2018){Henning}, {Sayre}, {Reichardt}, {Ade},
  {Anderson}, {Austermann}, {Beall}, {Bender}, {Benson}, {Bleem}, {Carlstrom},
  {Chang}, {Chiang}, {Cho}, {Citron}, {Corbett Moran}, {Crawford}, {Crites},
  {de Haan}, {Dobbs}, {Everett}, {Gallicchio}, {George}, {Gilbert},
  {Halverson}, {Harrington}, {Hilton}, {Holder}, {Holzapfel}, {Hoover}, {Hou},
  {Hrubes}, {Huang}, {Hubmayr}, {Irwin}, {Keisler}, {Knox}, {Lee}, {Leitch},
  {Li}, {Lowitz}, {Manzotti}, {McMahon}, {Meyer}, {Mocanu}, {Montgomery},
  {Nadolski}, {Natoli}, {Nibarger}, {Novosad}, {Padin}, {Pryke}, {Ruhl},
  {Saliwanchik}, {Schaffer}, {Sievers}, {Smecher}, {Stark}, {Story}, {Tucker},
  {Vanderlinde}, {Veach}, {Vieira}, {Wang}, {Whitehorn}, {Wu}, \&
  {Yefremenko}}]{henning/etal:2018}
{Henning}, J.~W., {Sayre}, J.~T., {Reichardt}, C.~L., {et~al.} 2018, \apj, 852,
  97

\bibitem[{{HI4PI Collaboration} {et~al.}(2016){HI4PI Collaboration}, {Ben
  Bekhti}, {Fl{\"o}er}, {Keller}, {Kerp}, {Lenz}, {Winkel}, {Bailin},
  {Calabretta}, {Dedes}, {Ford}, {Gibson}, {Haud}, {Janowiecki}, {Kalberla},
  {Lockman}, {McClure-Griffiths}, {Murphy}, {Nakanishi}, {Pisano}, \&
  {Staveley-Smith}}]{HI4PI:2016}
{HI4PI Collaboration}, {Ben Bekhti}, N., {Fl{\"o}er}, L., {et~al.} 2016, \aap,
  594, A116

\bibitem[{{Hildebrandt} {et~al.}(2017){Hildebrandt}, {Viola}, {Heymans},
  {Joudaki}, {Kuijken}, {Blake}, {Erben}, {Joachimi}, {Klaes}, {Miller},
  {Morrison}, {Nakajima}, {Verdoes Kleijn}, {Amon}, {Choi}, {Covone}, {de
  Jong}, {Dvornik}, {Fenech Conti}, {Grado}, {Harnois-D{\'e}raps}, {Herbonnet},
  {Hoekstra}, {K{\"o}hlinger}, {McFarland}, {Mead}, {Merten}, {Napolitano},
  {Peacock}, {Radovich}, {Schneider}, {Simon}, {Valentijn}, {van den Busch},
  {van Uitert}, \& {Van Waerbeke}}]{hildebrandt/etal:2017}
{Hildebrandt}, H., {Viola}, M., {Heymans}, C., {et~al.} 2017, \mnras, 465, 1454

\bibitem[{{Hivon} {et~al.}(2002){Hivon}, {G{\' o}rski}, {Netterfield}, {Crill},
  {Prunet}, \& {Hansen}}]{hivon/etal:2002}
{Hivon}, E., {G{\' o}rski}, K.~M., {Netterfield}, C.~B., {et~al.} 2002, \apj,
  567, 2

\bibitem[{Howlett {et~al.}(2012)Howlett, Lewis, Hall, \&
  Challinor}]{howlett/etal:2012}
Howlett, C., Lewis, A., Hall, A., \& Challinor, A. 2012, \jcap, 1204, 027

\bibitem[{{Hu}(2000)}]{hu:2000}
{Hu}, W. 2000, \prd, 62, 043007

\bibitem[{{Huang} {et~al.}(2018){Huang}, {Addison}, {Weiland}, \&
  {Bennett}}]{huang/etal:2018}
{Huang}, Y., {Addison}, G.~E., {Weiland}, J.~L., \& {Bennett}, C.~L. 2018,
  \apj, 869, 38

\bibitem[{Hunter(2007)}]{hunter:2007}
Hunter, J.~D. 2007, Computing in Science \& Engineering, 9, 90

\bibitem[{{Joudaki} {et~al.}(2017){Joudaki}, {Blake}, {Heymans}, {Choi},
  {Harnois-Deraps}, {Hildebrandt}, {Joachimi}, {Johnson}, {Mead}, {Parkinson},
  {Viola}, \& {van Waerbeke}}]{joudaki/etal:2016}
{Joudaki}, S., {Blake}, C., {Heymans}, C., {et~al.} 2017, \mnras, 465, 2033

\bibitem[{{Knox} \& {Millea}(2020)}]{knox/millea:2020}
{Knox}, L., \& {Millea}, M. 2020, \prd, 101, 043533

\bibitem[{{Lenz} {et~al.}(2019){Lenz}, {Dor{\'e}}, \&
  {Lagache}}]{lenz/dore/lagache:2019}
{Lenz}, D., {Dor{\'e}}, O., \& {Lagache}, G. 2019, \apj, 883, 75

\bibitem[{{Lewis}(2019)}]{getdist2019}
{Lewis}, A. 2019, arXiv e-prints, arXiv:1910.13970

\bibitem[{{Lewis} \& {Bridle}(2002)}]{lewis/bridle:2002}
{Lewis}, A., \& {Bridle}, S. 2002, \prd, 66, 103511

\bibitem[{{Lewis} \& {Challinor}(2006)}]{lewis/challinor:2006}
{Lewis}, A., \& {Challinor}, A. 2006, \physrep, 429, 1

\bibitem[{Lewis {et~al.}(2000)Lewis, Challinor, \&
  Lasenby}]{lewis/challinor/lasenby:2000}
Lewis, A., Challinor, A., \& Lasenby, A. 2000, \apj, 538, 473

\bibitem[{{Li} {et~al.}(2023){Li}, {Louis}, {Calabrese}, {Jense}, {Alonso},
  {Atkins}, {Bond}, {Choi}, {Dunkley}, {Fabbian}, {Garrido}, {H. Jaffe},
  {Madhavacheril}, {Meerburg}, {Natale}, \& {Qu}}]{li/etal:2023}
{Li}, Z., {Louis}, T., {Calabrese}, E., {et~al.} 2023, \jcap, 2023, 048

\bibitem[{{Madhavacheril} {et~al.}(2024){Madhavacheril}, {Qu}, {Sherwin},
  {MacCrann}, {Li}, {Abril-Cabezas}, {Ade}, {Aiola}, {Alford}, {Amiri},
  {Amodeo}, {An}, {Atkins}, {Austermann}, {Battaglia}, {Battistelli}, {Beall},
  {Bean}, {Beringue}, {Bhandarkar}, {Biermann}, {Bolliet}, {Bond}, {Cai},
  {Calabrese}, {Calafut}, {Capalbo}, {Carrero}, {Challinor}, {Chesmore}, {Cho},
  {Choi}, {Clark}, {C{\'o}rdova Rosado}, {Cothard}, {Coughlin}, {Coulton},
  {Crowley}, {Dalal}, {Darwish}, {Devlin}, {Dicker}, {Doze}, {Duell}, {Duff},
  {Duivenvoorden}, {Dunkley}, {D{\"u}nner}, {Fanfani}, {Fankhanel}, {Farren},
  {Ferraro}, {Freundt}, {Fuzia}, {Gallardo}, {Garrido}, {Givans}, {Gluscevic},
  {Golec}, {Guan}, {Hall}, {Halpern}, {Han}, {Harrison}, {Hasselfield},
  {Healy}, {Henderson}, {Hensley}, {Herv{\'\i}as-Caimapo}, {Hill}, {Hilton},
  {Hilton}, {Hincks}, {Hlo{\v{z}}ek}, {Ho}, {Huber}, {Hubmayr}, {Huffenberger},
  {Hughes}, {Irwin}, {Isopi}, {Jense}, {Keller}, {Kim}, {Knowles}, {Koopman},
  {Kosowsky}, {Kramer}, {Kusiak}, {La Posta}, {Lague}, {Lakey}, {Lee}, {Li},
  {Limon}, {Lokken}, {Louis}, {Lungu}, {MacInnis}, {Maldonado}, {Maldonado},
  {Mallaby-Kay}, {Marques}, {McMahon}, {Mehta}, {Menanteau}, {Moodley},
  {Morris}, {Mroczkowski}, {Naess}, {Namikawa}, {Nati}, {Newburgh}, {Nicola},
  {Niemack}, {Nolta}, {Orlowski-Scherer}, {Page}, {Pandey}, {Partridge},
  {Prince}, {Puddu}, {Radiconi}, {Robertson}, {Rojas}, {Sakuma}, {Salatino},
  {Schaan}, {Schmitt}, {Sehgal}, {Shaikh}, {Sierra}, {Sievers}, {Sif{\'o}n},
  {Simon}, {Sonka}, {Spergel}, {Staggs}, {Storer}, {Switzer}, {Tampier},
  {Thornton}, {Trac}, {Treu}, {Tucker}, {Ullom}, {Vale}, {Van Engelen}, {Van
  Lanen}, {van Marrewijk}, {Vargas}, {Vavagiakis}, {Wagoner}, {Wang}, {Wenzl},
  {Wollack}, {Xu}, {Zago}, \& {Zheng}}]{madhavacheril/etal:2024}
{Madhavacheril}, M.~S., {Qu}, F.~J., {Sherwin}, B.~D., {et~al.} 2024, \apj,
  962, 113

\bibitem[{{McCarthy} {et~al.}(2018){McCarthy}, {Bird}, {Schaye},
  {Harnois-Deraps}, {Font}, \& {van Waerbeke}}]{mccarthy/etal:2018}
{McCarthy}, I.~G., {Bird}, S., {Schaye}, J., {et~al.} 2018, \mnras, 476, 2999

\bibitem[{{Nguyen} {et~al.}(2023){Nguyen}, {Huterer}, \&
  {Wen}}]{nguyen/huterer/wen:2023}
{Nguyen}, N.-M., {Huterer}, D., \& {Wen}, Y. 2023, \prl, 131, 111001

\bibitem[{{Nicola} {et~al.}(2021){Nicola}, {Garc{\'\i}a-Garc{\'\i}a}, {Alonso},
  {Dunkley}, {Ferreira}, {Slosar}, \& {Spergel}}]{nicola/etal:2021}
{Nicola}, A., {Garc{\'\i}a-Garc{\'\i}a}, C., {Alonso}, D., {et~al.} 2021,
  \jcap, 2021, 067

\bibitem[{{\sorthelp{Planck Collaboration 2011A}}{Planck Collaboration
  I}(2011)}]{planck2011-1.1}
{\sorthelp{Planck Collaboration 2011A}}{Planck Collaboration I}. 2011, \aap,
  536, A1

\bibitem[{{\sorthelp{Planck Collaboration 2011R}}{Planck Collaboration
  XVIII}(2011)}]{planck2011-6.6}
{\sorthelp{Planck Collaboration 2011R}}{Planck Collaboration XVIII}. 2011,
  \aap, 536, A18

\bibitem[{{\sorthelp{Planck Collaboration 2014G}}{Planck Collaboration
  VII}(2014)}]{planck2013-p03c}
{\sorthelp{Planck Collaboration 2014G}}{Planck Collaboration VII}. 2014, \aap,
  571, A7

\bibitem[{{\sorthelp{Planck Collaboration 2014P}}{Planck Collaboration
  XVI}(2014)}]{planck2013-p11}
{\sorthelp{Planck Collaboration 2014P}}{Planck Collaboration XVI}. 2014, \aap,
  571, A16

\bibitem[{{\sorthelp{Planck Collaboration 2014W}}{Planck Collaboration
  XXIII}(2014)}]{planck2013-p09}
{\sorthelp{Planck Collaboration 2014W}}{Planck Collaboration XXIII}. 2014,
  \aap, 571, A23

\bibitem[{{\sorthelp{Planck Collaboration 2015I}}{Planck Collaboration
  IX}(2016)}]{planck2014-a11}
{\sorthelp{Planck Collaboration 2015I}}{Planck Collaboration IX}. 2016, \aap,
  594, A9

\bibitem[{{\sorthelp{Planck Collaboration 2015K}}{Planck Collaboration
  XI}(2016)}]{planck2014-a13}
{\sorthelp{Planck Collaboration 2015K}}{Planck Collaboration XI}. 2016, \aap,
  594, A11

\bibitem[{{\sorthelp{Planck Collaboration 2015L}}{Planck Collaboration
  XII}(2016)}]{planck2014-a14}
{\sorthelp{Planck Collaboration 2015L}}{Planck Collaboration XII}. 2016, \aap,
  594, A12

\bibitem[{{\sorthelp{Planck Collaboration 2018E}}{Planck Collaboration
  V}(2020)}]{planck2016-l05}
{\sorthelp{Planck Collaboration 2018E}}{Planck Collaboration V}. 2020, \aap,
  641, A5

\bibitem[{{\sorthelp{Planck Collaboration 2018F}}{Planck Collaboration
  VI}(2020)}]{planck2016-l06}
{\sorthelp{Planck Collaboration 2018F}}{Planck Collaboration VI}. 2020, \aap,
  641, A6

\bibitem[{{\sorthelp{Planck Collaboration 2018H}}{Planck Collaboration
  VIII}(2019)}]{planck2016-l08}
{\sorthelp{Planck Collaboration 2018H}}{Planck Collaboration VIII}. 2019, \aap,
  641, A8

\bibitem[{{\sorthelp{Planck Collaboration IntZZA}}{Planck Collaboration Int.
  LI}(2017)}]{planck2016-LI}
{\sorthelp{Planck Collaboration IntZZA}}{Planck Collaboration Int. LI}. 2017,
  \aap, 607, A95

\bibitem[{{\sorthelp{Planck Collaboration IntZZG}}{Planck Collaboration Int.
  LVII}(2020)}]{planck2020-LVII}
{\sorthelp{Planck Collaboration IntZZG}}{Planck Collaboration Int. LVII}. 2020,
  \aap, 643, A42

\bibitem[{{Reichardt} {et~al.}(2012){Reichardt}, {Shaw}, {Zahn}, {Aird},
  {Benson}, {Bleem}, {Carlstrom}, {Chang}, {Cho}, {Crawford}, {Crites}, {de
  Haan}, {Dobbs}, {Dudley}, {George}, {Halverson}, {Holder}, {Holzapfel},
  {Hoover}, {Hou}, {Hrubes}, {Joy}, {Keisler}, {Knox}, {Lee}, {Leitch},
  {Lueker}, {Luong-Van}, {McMahon}, {Mehl}, {Meyer}, {Millea}, {Mohr},
  {Montroy}, {Natoli}, {Padin}, {Plagge}, {Pryke}, {Ruhl}, {Schaffer},
  {Shirokoff}, {Spieler}, {Staniszewski}, {Stark}, {Story}, {van Engelen},
  {Vanderlinde}, {Vieira}, \& {Williamson}}]{reichardt/etal:2012}
{Reichardt}, C.~L., {Shaw}, L., {Zahn}, O., {et~al.} 2012, \apj, 755, 70

\bibitem[{{Riess} {et~al.}(2022){Riess}, {Yuan}, {Macri}, {Scolnic}, {Brout},
  {Casertano}, {Jones}, {Murakami}, {Anand}, {Breuval}, {Brink}, {Filippenko},
  {Hoffmann}, {Jha}, {D'arcy Kenworthy}, {Mackenty}, {Stahl}, \&
  {Zheng}}]{riess/etal:2022}
{Riess}, A.~G., {Yuan}, W., {Macri}, L.~M., {et~al.} 2022, \apjl, 934, L7

\bibitem[{{Rosenberg} {et~al.}(2022){Rosenberg}, {Gratton}, \&
  {Efstathiou}}]{rosenberg/gratton/efstathiou:2022}
{Rosenberg}, E., {Gratton}, S., \& {Efstathiou}, G. 2022, \mnras, 517, 4620

\bibitem[{{Schwarz} {et~al.}(2016){Schwarz}, {Copi}, {Huterer}, \&
  {Starkman}}]{schwarz/etal:2016}
{Schwarz}, D.~J., {Copi}, C.~J., {Huterer}, D., \& {Starkman}, G.~D. 2016,
  Classical and Quantum Gravity, 33, 184001

\bibitem[{{Sellentin} \& {Starck}(2019)}]{sellentin/starck:2019}
{Sellentin}, E., \& {Starck}, J.-L. 2019, \jcap, 2019, 021

\bibitem[{{Smith} {et~al.}(2022){Smith}, {Lucca}, {Poulin}, {Abellan},
  {Balkenhol}, {Benabed}, {Galli}, \& {Murgia}}]{smith/etal:2022}
{Smith}, T.~L., {Lucca}, M., {Poulin}, V., {et~al.} 2022, \prd, 106, 043526

\bibitem[{{Story} {et~al.}(2013){Story}, {Reichardt}, {Hou}, {Keisler}, {Aird},
  {Benson}, {Bleem}, {Carlstrom}, {Chang}, {Cho}, {Crawford}, {Crites}, {de
  Haan}, {Dobbs}, {Dudley}, {Follin}, {George}, {Halverson}, {Holder},
  {Holzapfel}, {Hoover}, {Hrubes}, {Joy}, {Knox}, {Lee}, {Leitch}, {Lueker},
  {Luong-Van}, {McMahon}, {Mehl}, {Meyer}, {Millea}, {Mohr}, {Montroy},
  {Padin}, {Plagge}, {Pryke}, {Ruhl}, {Sayre}, {Schaffer}, {Shaw}, {Shirokoff},
  {Spieler}, {Staniszewski}, {Stark}, {van Engelen}, {Vanderlinde}, {Vieira},
  {Williamson}, \& {Zahn}}]{story/etal:2013}
{Story}, K.~T., {Reichardt}, C.~L., {Hou}, Z., {et~al.} 2013, \apj, 779, 86

\bibitem[{{Tristram} {et~al.}(2021){Tristram}, {Banday}, {G{\'o}rski},
  {Keskitalo}, {Lawrence}, {Andersen}, {Barreiro}, {Borrill}, {Eriksen},
  {Fernandez-Cobos}, {Kisner}, {Mart{\'\i}nez-Gonz{\'a}lez}, {Partridge},
  {Scott}, {Svalheim}, {Thommesen}, \& {Wehus}}]{tristram/etal:2021}
{Tristram}, M., {Banday}, A.~J., {G{\'o}rski}, K.~M., {et~al.} 2021, \aap, 647,
  A128

\bibitem[{{Tristram} {et~al.}(2024){Tristram}, {Banday}, {Douspis}, {Garrido},
  {G{\'o}rski}, {Henrot-Versill{\'e}}, {Hergt}, {Ili{\'c}}, {Keskitalo},
  {Lagache}, {Lawrence}, {Partridge}, \& {Scott}}]{tristram/etal:2024}
{Tristram}, M., {Banday}, A.~J., {Douspis}, M., {et~al.} 2024, \aap, 682, A37

\bibitem[{{van Engelen} {et~al.}(2012){van Engelen}, {Keisler}, {Zahn}, {Aird},
  {Benson}, {Bleem}, {Carlstrom}, {Chang}, {Cho}, {Crawford}, {Crites}, {de
  Haan}, {Dobbs}, {Dudley}, {George}, {Halverson}, {Holder}, {Holzapfel},
  {Hoover}, {Hou}, {Hrubes}, {Joy}, {Knox}, {Lee}, {Leitch}, {Lueker},
  {Luong-Van}, {McMahon}, {Mehl}, {Meyer}, {Millea}, {Mohr}, {Montroy},
  {Natoli}, {Padin}, {Plagge}, {Pryke}, {Reichardt}, {Ruhl}, {Sayre},
  {Schaffer}, {Shaw}, {Shirokoff}, {Spieler}, {Staniszewski}, {Stark}, {Story},
  {Vanderlinde}, {Vieira}, \& {Williamson}}]{vanengelen/etal:2012}
{van Engelen}, A., {Keisler}, R., {Zahn}, O., {et~al.} 2012, \apj, 756, 142

\bibitem[{Virtanen {et~al.}(2020)Virtanen, Gommers, Oliphant, Haberland, Reddy,
  Cournapeau, Burovski, Peterson, Weckesser, Bright, {van der Walt}, Brett,
  Wilson, Millman, Mayorov, Nelson, Jones, Kern, Larson, Carey, Polat, Feng,
  Moore, {VanderPlas}, Laxalde, Perktold, Cimrman, Henriksen, Quintero, Harris,
  Archibald, Ribeiro, Pedregosa, {van Mulbregt}, \& {SciPy 1.0
  Contributors}}]{virtanen/etal:2020}
Virtanen, P., Gommers, R., Oliphant, T.~E., {et~al.} 2020, Nature Methods, 17,
  261

\end{thebibliography}

\appendix

\section{Corrections to Noise Model from Full Focal Plane Simulations}

\label{app:noise}

Here we describe in more detail the heuristic correction applied to the noise power measured from the Planck FFP10 simulations to better match the noise spectra from the HM data maps at 100, 143, and 217~GHz. For these calculations we combined the missing pixels across the two HM for each band, since an identical mask is required to accurately remove the sky signal in a half-mission difference (see below). 

As mentioned in Section~\ref{sec:noise}, we use the mean of the 300 FFP10 noise simulations to estimate $\psi_{\ell}^{\nu,\textrm{HM}}$, the harmonic-space rescaling factor applied to the white noise power estimated from the pixel variance values supplied with the Planck maps. We then perform a leading-order correction for the mismatch with the noise in the real data by rescaling $\psi_{\ell}^{\nu,\textrm{HM}}$ by a fourth-order polynomial in $\ell$:
\be
\psi_{\ell}^{\nu,\textrm{HM}}\rightarrow \psi_{\ell}^{\nu,\textrm{HM}}\left(\sum_{n=0}^4\alpha_n^{\nu}\left(\ell/500\right)^n\right)
\ee
The noise power in the data can be estimated from differencing HM auto and cross-spectra \citep{li/etal:2023}, however this leads to large scatter at low multipoles due to chance signal-noise correlations with the high-amplitude CMB modes. To remove this effect and null the sky signal completely we computed power spectra of the HM difference maps for each band. These spectra are only sensitive to the sum of the HM noise spectra, however we find that the polynomial correction factors derived from scaling the FFP simulation power to match these spectra also adequately improve agreement to the high multipoles of the auto-cross difference for each HM separately. The polynomial coefficients were derived by minimizing a simple sum-of-square residuals of the form
\be
S^2=\sum_{\ell=20}^{\ell_{\rm max}}\frac{2\ell+1}{\left(\tilde{N}_{\ell}^{\nu,\textrm{HM-diff model}}\right)^2}\left(\tilde{N}_{\ell}^{\nu,\textrm{HM-diff model}}-\tilde{N}_{\ell}^{\nu,\textrm{HM-diff data}}\right)^2,
\label{eqn:noise}
\ee
where the tildes denote pseudo-spectra and the maximum multipole was set to 1400 for 100~GHz and 2700 for 143 and 217~GHz. These values were conservatively chosen to exceed the maximum multipoles used for the cosmological fitting by at least 200. The prefactor in equation~\ref{eqn:noise} is an approximate inverse-variance weight. We neglect mode-coupling from the mask here but reiterate that the details of this noise model correction do not have a significant impact on the $A_L$ parameter results.

\section{Validating Likelihood with Simulations}
\label{app:sims}

In this Appendix we describe simulations we performed to validate the power spectrum covariance matrix estimation described in Section~\ref{sec:computations}. The Planck Collaboration already performed a range of validation tests for the cosmological results (see PL15 and PL18). Our goal here is not to repeat these tests but to check aspects of our analysis that differ from the Planck papers, specifically (1) the validity of the Improved Narrow-Kernel Approximation (Section~\ref{sec:NKA}), and (2) the overall accuracy of the covariance matrix approximation as additional masking is applied.

We generated 500 data-like simulation realizations. We simulated 100, 143, and 217~GHz HM maps containing CMB, foreground, and noise fluctuations and then applied masking and estimated power spectra as described in Section~\ref{sec:pseudoCl}. The CMB and foregrounds were generated as statistically isotropic and Gaussian fields using the best-fit parameters from the baseline Planck 2018 \lcdm\ fit including all temperature, polarization and lensing data. Clearly, the Galactic emission is not statistically isotropic, but it is approximated as such for the purposes of the covariance matrix estimation in the Planck analyses, and since we follow this approximation in our calculations we did not construct these simulations to test it.\footnote[21]{We did separately test the impact of changing the Galactic foreground levels assumed in the covariance matrix construction for different masks. This test is described in Appendix~\ref{app:dusttemplates}.} For the noise we first generated anisotropic white noise maps, using the pixel variance maps provided with each Planck HM data map. To simulate the non-white, scale dependent noise observed in the real Planck data we then applied a harmonic space rescaling of the noise fluctuations such that the shape of the noise power matches the $\psi_{\ell}^{\nu, \textrm{HM}}$ functions described in Section~\ref{sec:noise} and Appendix~\ref{app:noise}.

\begin{figure*}
\centering
\includegraphics[]{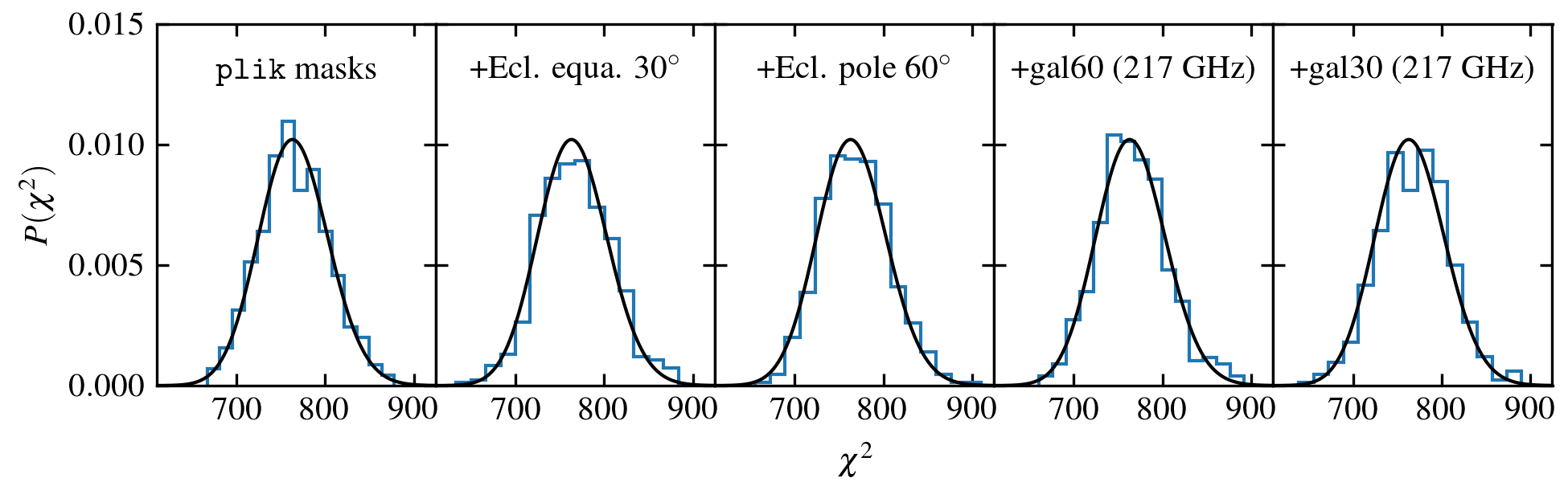}
\caption{Distribution of power spectrum $\chi^2$ values across 500 simulation realizations for five sets of masks used in this work (blue histograms), compared to the expected $\chi^2$ probability density (black lines). The simulated data vector for each realization and set of masks consists of 765 power spectrum bandpowers from 100, 143, $143\times217$ and 217~GHz, matching the analysis of the real Planck data. The $\chi^2$ values are computed against the input CMB-plus-foreground model used to generate the simulations. The first panel shows results from the original Planck \texttt{plik} likelihood masks (with very minor adjustment for additional bad pixels, see Section~\ref{sec:maps_masks}). The other panels show results from additional masking described in Sections~\ref{sec:ecliptic} and \ref{sec:dust}. Consistency of the histograms with the expected $\chi^2$ distribution tests the Improved Narrow-Kernel Approximation for estimating the covariance matrix (Section~\ref{sec:NKA}). The second through fifth panels demonstrate that additional sky masking does not significantly bias the covariance matrix calculation, even for cases where the sky fraction retained after masking is as low as $f_{\rm sky}\simeq0.2$ at 217~GHz.}
\label{fig:chi2_sims}
\end{figure*}

A direct and effective diagnostic of the power spectrum covariance accuracy is the $\chi^2$ distribution across simulation realizations against the known input. Figure~\ref{fig:chi2_sims} shows that the $\chi^2$ distributions are well recovered even for the more extreme masks used in this work (i.e., removing the most sky). With 500 realizations we can limit the $\chi^2$ bias from inaccuracies in our covariance matrix calculation to $\Delta\chi^2\lesssim5$. This is a meaningful limit based on tests where we deliberately modify the covariance matrix to create mismatches with the simulations. Introducing such mismatches in the modeling of the non-white noise correction, or the choice of the $100\times143$ and $100\times217$ fiducial spectra, for example, measurably bias the simulation $\chi^2$ (i.e, induce $\Delta\chi^2>5$), even though their impact on $A_L$ and other parameters is negligible ($\leq0.1\sigma$).

\section{Testing Spatial Dependence of Galactic Dust Templates}
\label{app:dusttemplates}

In this Appendix we describe two tests of the foreground modeling at 217~GHz for different masks. We focus on 217~GHz because $A_L$ constraints from the 217~GHz data show the strongest mask dependence, and because the dust foreground is significantly brighter at 217~GHz than at 100 or 143~GHz.

First, given the anisotropic nature of the Galactic emission, a natural question is whether the Galactic dust template shapes in $\ell$ in the cosmological fitting need to be modified as we change the masks. Here we describe a leading-order calculation we performed to test this. We follow a similar approach to that described for constructing the original \texttt{plik} dust templates, using 545~GHz as a sensitive tracer of the Galactic dust, and assuming that the 217~GHz dust fluctuations largely trace those at 545~GHz (Section~3.3.2 of PL18). In practice we did not directly modify the templates used in the likelihood code, but performed an equivalent modification to the data vectors.

We computed 545~GHz HM cross-spectra for the \texttt{plik} 217~GHz HM masks plus the modified masks from Sections~3 and 4, and applied the appropriate mask deconvolution matrices. For each modified mask we then fit a simple parametric model, a power law with free amplitude and index, plus a constant, to the modified mask / \texttt{plik} mask difference spectrum. We finally obtained a 217~GHz amplitude by matching this parametric model shape in $\ell$ against the corresponding modified mask / \texttt{plik} mask difference spectrum at 217~GHz. The product of this amplitude and the parametric model shape was then added to the 217~GHz power spectrum data file. Note that constraining the shape of the template modification using the 217~GHz difference spectra alone (without higher frequency data) is not possible due to large scatter from the CMB fluctuations in the difference spectra between pairs of masks.

For the ecliptic masks described in Section~3 (\texttt{plik}+\texttt{equa30}, \texttt{plik}+\texttt{pole60}, etc.), and for the combinations of the \texttt{plik} mask with the \texttt{gal60\_857} and \texttt{gal50\_857} masks described in Section~4, this procedure had a negligible impact compared to simply re-fitting the original Planck foreground model on each modified mask. The values of $A_L$ and other cosmological parameters shift at the $0.1\sigma$ level or less. For the more extreme dust masks used in Section~4, specifically, the \texttt{gal40\_857} and \texttt{gal30\_857} masks that remove even more of the sky based on 857~GHz intensity, the effect is larger, $0.2-0.3\sigma$ in $A_L$. These shifts do not qualitatively impact the trends in $A_L$ with additional masking discussed in Section~4.

We note that the procedure outlined above is approximate, for example the power-law-plus-constant model is not a perfect match for all the 545~GHz modified mask / original mask difference spectra. Additionally, we did not compute full covariances for the 545 and 217~GHz difference spectra to determine the template parameters and 217~GHz amplitudes, but instead performed simplified diagonal least-squares fits. Given the small cosmological impact of this leading-order correction we did not investigate more accurate corrections.

Second, we investigated updating the fiducial 217~GHz foreground model assumed for the power spectrum covariance matrix calculation, replacing the original foreground model (obtained using the original \texttt{plik} masks) with their counterparts from fits to the modified masks. In principle we could use an iterative approach (update covariance matrix, refit, update again, etc.), however in practice we found that this modification made negligible difference to parameters ($\leq0.1\sigma$ in $A_L$, similarly small shifts in uncertainties) for all cases. In general, an inaccurate covariance matrix can bias posterior constraints in addition to overestimating or underestimating the posterior widths \citep[e.g.,][]{sellentin/starck:2019}. The reason this is not happening in this case is essentially just that the foreground power is subdominant to the CMB sample variance as an uncertainty source (see Figure~32 of PL18). Even a significant change in the foreground power, at the level of tens of per cent, only impacts the overall covariance at the few per cent level. Note that the power difference in Figure~\ref{fig:ecl_power_diff_217} corresponds to roughly 20\% of the total 217~GHz foreground model on the original \texttt{plik} masks. At sufficiently high multipoles the foregrounds overtake the CMB as a power source, however the spectra at these scales ($\ell>2000$ for 217~GHz) also become noisy and contribute little to the overall cosmological constraints.

\end{document}